\pdfoutput=1

\documentclass[12pt]{article}
\usepackage{amsmath,amssymb,latexsym,cite,eucal,amsthm,enumerate}
\usepackage[matrix,arrow,frame,import,curve,color]{xy}
\usepackage{tikz}
\usetikzlibrary{arrows}
\usetikzlibrary{decorations.pathmorphing}
\usetikzlibrary{decorations.markings}
\usetikzlibrary{calc}

\usepackage{graphicx}

\newtheorem{theorem}{Theorem}
\newtheorem{prop}[theorem]{Proposition}

\newtheorem{conjecture}[theorem]{Conjecture}

\theoremstyle{definition}
\newtheorem{definition}[theorem]{Definition}

\newif\iffigs\figstrue

\DeclareFontFamily{U}{rsf}{}
\DeclareFontShape{U}{rsf}{m}{n}{
  <5> <6> rsfs5 <7> <8> <9> rsfs7 <10-> rsfs10}{}
\DeclareMathAlphabet\Scr{U}{rsf}{m}{n}

\def\O{\Scr{O}}
\def\C{{\mathbb C}}
\def\P{{\mathbb P}}

\def\Q{{\mathbb Q}}
\def\R{{\mathbb R}}
\def\Z{{\mathbb Z}}

\def\la{{\langle}}
\def\ra{{\rangle}}

\def\Hom{\operatorname{Hom}}

\def\Spec{\operatorname{Spec}}

\def\Conv{\operatorname{Conv}}
\def\Crit{\operatorname{Crit}}

\def\SU{\operatorname{SU}}
\def\GU{\operatorname{U{}}}

\def\rank{\operatorname{rank}}

\def\Cone{\operatorname{Cone}}

\def\frak{\mathfrak}

\def\p{\partial}

\def\CY{Calabi--Yau}
\def\LG{Landau--Ginzburg}

\def\GLSM{gauged linear $\sigma$-model}
\def\NLSM{non-linear $\sigma$-model}
\def\SCFT{superconformal field theory}
\def\SCFTS{superconformal field theories}
\def\FI{Fayet--Iliopoulos}

\def\cM{{\Scr M}}

\def\cN{{\Scr N}}

\def\cA{{\Scr A}}
\def\cB{{\Scr B}}

\def\DC{\mathbf{D}^b}

\def\QED{$\quad\blacksquare$}

\def\ff#1#2{{\textstyle\frac{#1}{#2}}}

\def\eqn#1#2{\begin{equation}#2
  \ifx{#1}{}\else\label{#1}\fi\end{equation}}

\begin{document}

\begin{titlepage}
\begin{flushright}
June 2015
\end{flushright}
\vspace{.5cm}
\begin{center}
\baselineskip=16pt
{\fontfamily{ptm}\selectfont\bfseries\huge
General Mirror Pairs for\\Gauged Linear Sigma Models\\[20mm]}
{\bf\large Paul S.~Aspinwall and M.~Ronen Plesser
 } \\[7mm]

{\small

Departments of Mathematics and Physics,\\ 
  Box 90320, \\ Duke University,\\ 
 Durham, NC 27708-0320 \\ \vspace{6pt}

 }

\end{center}

\begin{center}
{\bf Abstract}
\end{center}

We carefully analyze the conditions for an abelian \GLSM\ to exhibit
nontrivial IR behavior described by a nonsingular \SCFT\ 
determining a superstring vacuum.
This is done
without reference to a geometric phase, by associating singular
behavior to a noncompact space of (semi-)classical vacua.
We find that models determined by {\em reflexive\/} combinatorial data are
nonsingular for generic values of their parameters.  This condition
has the pleasant feature that the mirror of a nonsingular \GLSM\ is
another such model, but it is clearly too strong and we provide an
example of a non-reflexive mirror pair.  We discuss a weaker condition
inspired by considering extremal transitions, which is also mirror
symmetric and which we conjecture to be sufficient.  We apply these
ideas to extremal transitions and to understanding the way in which
both Berglund--H\"ubsch mirror symmetry and the Vafa--Witten mirror
orbifold with discrete torsion can be seen as special cases of the
general combinatorial duality of \GLSM s.  In the former case we
encounter an example showing that our weaker condition is still not
necessary.



\end{titlepage}

\vfil\break


\section{Introduction}    \label{s:intro}

Witten's abelian \GLSM  \cite{W:phase} has been a powerful tool in mirror
symmetry for many years \cite{AGM:II,MP:inst}.   For a \CY\ space $X$
determined as a complete intersection in a toric variety, the \GLSM\
provides a UV safe model whose IR behavior is governed by the \SCFT\
determined by the \NLSM\ with target space $X$.  The combinatorial
duality of Batyrev and Borisov \cite{Bat:m,Boris:m} relating mirror
families can be stated \cite{Morrison:1995yh, Hori:2000kt} as a
duality relating \GLSM s with identical IR physics, including an
explicit mapping of the parameters.  


Conventionally the focus has been on the {\em geometry\/} of mirror
symmetry. The desired \CY\ $X$ is a subvariety of a toric variety,
which is assumed to be Gorenstein. However, both from a physics and a
mathematical point of view, it is necessary to widen one's perspective
from the geometry of varieties, as has long been known \cite{AG:gmi}. 

From a physics perspective, mirror symmetry relates isomorphic
$N=(2,2)$ superconformal field theories with some fixed
central charge $3d$ and integral $U(1)_R$ charges.
Such an object may, or may not, arise as a nonlinear
$\sigma$-model on a \CY\ $d$-fold. In either case, it is equally
useful for compactifying a string model. Similarly, the mirror of a
\CY\ $d$-fold may, or may not, be another \CY\ $d$-fold.  In the
context of \GLSM s a given model may, or may not, exhibit a geometric
phase corresponding to a large-radius \CY\ $d$-fold.   In general, it
is probable (although currently no examples are known) that a \GLSM\ can have no
geometric description {\em anywhere\/} in its parameter space.\footnote{We
  know models that do not have a geometric phase but in all known
  cases there is a mirror model that does exhibit such a phase.}

From the mathematical perspective, mirror symmetry is an equivalence
between two triangulated categories (or dg-categories). The
essentially structure of a \CY\ $d$-fold is thus its underlying \CY\
category structure. Now, an abstract \CY\ category of dimension $d$
may, or may not, be expressible as the derived category of some actual
\CY\ $d$-fold. This is really the exact same statement as the above
paragraph. Thus, in this paper we take the perspective that one is
primarily interested in generating $N=(2,2)$ superconformal field
theories with spectral flow, or equivalently, \CY\ categories. Whether
or not such a model exhibits a geometric limit in any way is a
difficult question to answer and will be of secondary importance to us.

The main focus of this paper is a precise condition for the
nonsingularity of a \GLSM\ without reference to a geometric phase. We
try to be rigorous and complete. The data provided is a set of chiral
fields, an abelian gauge group (which may have discrete factors) and a
superpotential. Out of these data come two pointsets $\cA$ and $\cB$
that live in dual lattices.  We demand that the theory defined by
these pointsets be nonsingular for {\em generic\/} values of the
parameters. Not surprisingly, in theorem \ref{th:main}, we recover
reflexivity of the pointsets as a sufficient condition. However, it is
clear that reflexivity is not {\em necessary}.

Geometrically an extremal transition (or generalized conifold
transition) 
connects two topologically distinct models by a
blow-down followed by a deformation. It has long been realized that
this is easy to implement in terms of the \GLSM\ data \cite{Avram:1997rs}
by removing points from $\cA$ to shrink its convex hull and then
adding points to $\cB$ to increase its convex hull. The same mechanism
works whether there is a geometrical interpretation or not and we will
still refer to it as an extremal transition.  In the four-dimensional
spacetime theory (for $d=3$) these are Coulomb--Higgs transitions that
can be analyzed in quantum field theory in a limit in which gravity decouples.
In this paper we will use such transitions to motivate a less
restrictive condition for nonsingularity of the IR physics and the
associated extension of the mirror duality given by 
conjecture \ref{conj:main}.


The extremal transitions between models with geometric phases and
those without are quite interesting and we give some examples. These
will also connect reflexive pairs with nonreflexive pairs thus showing
that one must drop the reflexive condition to fully understand the
connected web of compactifications.

In section \ref{s:glsm} we define the data that goes into constructing
a \GLSM. Building on the work of \cite{Morrison:1995yh} we see that
the question of singularities is very closely tied with Gelfand,
Kapranov and Zelevinski's work on discriminants \cite{GKZ:book}. In
particular, for both the chiral 
and 
twisted chiral moduli spaces
we have a combinatorial construction of
the set of singular theories based
on faces of convex hulls of pointsets. This allows us to formulate
theorem \ref{th:main} which states sufficient (but not necessary)
conditions for a model to be nonsingular for generic moduli.

We give examples of consistent pointsets $\cA$ and $\cB$ in section
\ref{s:ex}. In particular, we give a mirror pair that does not consist
of a reflexive pair. Dropping points on the boundary on the convex
hull of these pointsets to violate the conditions of theorem
\ref{th:main} leads to extremal transitions. We discuss some
interesting extremal transitions from models with a geometrical
description to ones without. In particular we see how non-geometrical
Gepner models fit into the connected web of $N=(2,2)$ superconformal
field theories.

In section \ref{s:int} we consider more examples which are simplified
by dropping points interior to the convex hull. This gives a simple
description of Berglund and H\"ubsch's mirror symmetry construction
\cite{Berglund:1991pp}. We then reinterpret this example as a
nonreflexive model which violates the conditions of the conjecture and
yet remains nonsingular. This proves that even our weakened conditions
in the conjecture are still too strong. Finally we describe Vafa and
Witten's discrete torsion mirror pair \cite{VW:tor}.


\section{The \GLSM.}  \label{s:glsm}

\subsection{The pointsets $\cA$ and $\cB$.}

\def\fkg{\frak{g}}
Let us consider a general setting for \GLSM\ and describe how the data
describing the model can be written in the language of toric
geometry. We base this on the construction in \cite{Morrison:1995yh}.

The data determining our family of $N=(2,2)$ two-dimensional models is:
\begin{itemize}
\item A set of $n$ chiral superfields $\{x_1,\ldots, x_n\}$.
\item An abelian gauge group $G$ acting effectively and diagonally on
  $\{x_1,\ldots, x_n\}$.
\item A family of worldsheet superpotentials $W$, which are polynomials in
  $\{x_1,\ldots, x_n\}$ invariant under $G$.
\end{itemize}
For consistency of notation with the toric construction, we will use
$G$ to denote the {\em complexified\/} gauge group. That is, $G$ is of the
form $(\C^*)^r\times\Gamma$ for some finite group $\Gamma$.

We can present $G$ and its action on $\{x_1,\ldots, x_n\}$ in the
usual toric language as follows. First introduce the character group
of $G$:
\begin{equation}
  \widehat G = \Hom_{\Z}(G,\C^*),
\end{equation}
which is isomorphic to $\Z^r\oplus\Gamma$. Note that if $\fkg$ is the
Lie algebra of $G$ then
\begin{equation}
  \widehat G\otimes\C = \fkg^*.
\end{equation}
The diagonal $G$-action on $\{x_1,\ldots, x_n\}$ gives an obvious
eigenvalue map $q:G\to(\C^*)^n$. If the action is effective, $q$ is
injective. Applying $\Hom(-,\C^*)$ to $q$ gives a presentation of
$\widehat G$:
\begin{equation}
\xymatrix@1{
0\ar[r]&M\ar[r]^{A^t}&\Z^n\ar[r]^Q&\widehat G\ar[r]&0,
} \label{eq:kerQ}
\end{equation}
where $M$ is a lattice of rank $d=n-r$ and $A$ is a $d\times n$ matrix.
The map from $\Z^n$ to the free part of $\widehat G$ is given by the charge
matrix for the $(\C^*)^r$-action.

Let $N$ denote the dual lattice of $M$ and let $T_N=N\otimes_\Z\C^*$.
The matrix $A$ can be viewed
as having columns giving the coordinates of $n$ points in $N$. In this
way we identify the superfields $\{x_1,\ldots, x_n\}$ with
$n$ points in $N$. Let $\cA$ denote this pointset.
Thus from (\ref{eq:kerQ}) 
we now have the short exact sequence familiar from toric geometry
\cite{CLS:ToricVar} 
\begin{equation}
\xymatrix@1{
1\ar[r]&G\ar[r]^-{Q^t}&(\C^*)^{\cA} \ar[r]^-A&T_N\ar[r]&1.
}
\end{equation}

Now consider the superpotential. Suppose there are $m$ terms in this
polynomial. Each monomial represents a vector in $\Z^n$ given by the
list of powers in the monomial. Thus the powers of in the monomials in the
superpotential define a matrix giving a map
\begin{equation}
  P:\Z^m\to\Z^n.\label{eq:un}
\end{equation}
The superpotential is invariant under the $G$-action, which amounts to
the statement $QP=0$. The map $P$ must therefore factor
through the kernel of $Q$ given by (\ref{eq:kerQ}). This defines a map
\begin{equation}
  B:\Z^m\to M,
\end{equation}
and the map in (\ref{eq:un}) is given by the composition $P=A^t B$.

The columns of the $d\times m$ matrix $B$ give a set of $m$ points
in the lattice $M$. Let $\cB$ denote this pointset.  The condition
that $W$ is polynomial implies
\begin{equation}
\cB\subset\Cone(\Conv\cA)^\vee .
\end{equation}

Each element $\beta\in\cB$ determines a monomial, which we denote
\begin{equation}\label{eq:mmono}
  x^\beta = \prod_{\alpha\in\cA} x_\alpha^{\langle
    \beta,\alpha\rangle},
\end{equation}
and our family of superpotentials is
\begin{equation}
  W = \sum_{\beta\in\cB} b_\beta x^\beta,
\end{equation}
where $b_\beta\in\C$.

\subsection{The hyperplane conditions}

\def\ones#1{\mathbf{1}_{#1}}


The combinatorial data determine a family of \GLSM s.  In
general, these will exhibit trivial IR dynamics.   We
are interested in using the IR limit of such theories to construct
\SCFTS\ with integral $R$ charges.  In this section we determine the
conditions on the data under which this is the case.  In models with a
geometric interpretation this is equivalent to the \CY\ condition.  A
$(2,2)$ \SCFT\ exhibits a chiral $U(1)_R$ symmetry that is part of the
superconformal algebra.   Following \cite{W:phase} we assume
that a \GLSM\ will determine a nontrivial
\SCFT\ in the IR if it exhibits a chiral $R$-symmetry that is
non-anomalous, hence preserved along the RG flow.  

The eigenvalues of an $R$-symmetry which acts diagonally on the
$x_\alpha$'s give a map $R:\Q\to\Q^\cA$. Define $\nu=AR$ to get the
diagram
\begin{equation}
\xymatrix{
&&\Q^\cB\\
G\otimes_{\C^*}\Q\ar[r]&\Q^{\cA} \ar[r]^-A\ar[ur]^-{P^t}&
   N\otimes_{\Z}\Q\ar[r]\ar[u]^{B^t}&0\\
&\Q\ar[u]^R\ar[ur]^-\nu
}
\end{equation}
An unbroken $R$-symmetry will exist in our model if we can assign
rational charges to the chiral superfields such that each term in the
superpotential has charge one. That means 
\begin{equation}
 B^t\nu=P^t R =\ones{m},
\end{equation}
where $\ones{m}$ is a column vector of $m$ ones.
In other words, if we use $\nu$ to denote the image $\nu(1)$ in
$N\otimes\Q$, we have
\begin{equation}
  \langle \beta,\nu\rangle=1, \quad\forall \beta\in\cB.  \label{eq:Bplane}
\end{equation}

Our interest here is in \SCFTS\ suitable for constructing string vacua
with spacetime supersymmetry.  This imposes an additional condition,
namely that the $R$ charges of all gauge invariant chiral operators be
integral.  Such operators are associated by (\ref{eq:mmono}) to points
in $M$.
Thus $\nu\in N$.

Now we turn our attention to a similar condition for $\cA$.
The axial R-symmetry will be non-anomalous if for each generator of the {\em
  continuous\/} part $(\C^*)^r\subset G$ the sum of the charges of the chiral
fields vanishes \cite{W:phase}. 

If the finite group $\Gamma$ is nontrivial, we again have an
additional restriction from the requirement that the $R$ charges are
integral.  In particular, we consider the requirement that the
right-moving $R$ charge of states in sectors twisted by elements of
$\Gamma$ are integral. The argument closely follows a similar
construction in \cite{meMP:singlets}.  Consider an element $g \in
\Gamma$ of order $N$ acting as
\begin{equation}
x_\alpha\mapsto e^{\frac{2\pi iw_\alpha}N} x_\alpha
\end{equation}
Working in the extreme UV we can describe the vacuum in the NS sector
twisted by $g$ using free fields.  The right-moving $R$ charge can be
written in terms of the free field currents $\overline{\psi}_\alpha\psi_\alpha$ and
$\overline{\gamma}_\alpha\gamma_\alpha$ acting on the right- and
left-moving fermions in the chiral multiplet $x_\alpha$
\begin{equation}
R_{\textrm{right}}=\sum_\alpha (R_\alpha-1)\overline{\psi}_\alpha\psi_\alpha 
 + R_\alpha \overline{\gamma}_\alpha\gamma_\alpha.
\end{equation}
(These symmetries are of course separately broken by the interactions.)
In the NS sector twisted by $g$ the vacuum carries charge
$-w_\alpha/N$ (mod 1) under $\overline{\psi}_\alpha\psi_\alpha$ and
charge $w_\alpha/N$ (mod 1) under
$\overline{\gamma}_\alpha\gamma_\alpha$.  Inserting this we see that
the $R$ charge of the vacuum (and hence of the excitations above it by
gauge-invariant fields with integral $R$ charge) is
\begin{equation}
\langle g|R_{\textrm{right}}|g\rangle = \sum_\alpha (w_\alpha/N) \mod 1.  \label{eq:Gtors}
\end{equation}

Together with the requirement that the anomaly vanishes,
(\ref{eq:Gtors}) is equivalent to the statement that 
the monomial $\prod_{\alpha\in\cA} x_\alpha$ is
invariant under all of $G$. (This monomial may, or may not, be in the
superpotential.)

This can also be phrased as the condition $Q\ones{n}=0$.  We therefore
have a diagram inducing a map $\mu$:
\begin{equation}
\xymatrix{
0\ar[r]&M\ar[r]^{A^t}&\Z^{\cA}\ar[r]^Q&\widehat G\ar[r]&0\\
&&\Z\ar@{.>}[ul]^{\mu}\ar[u]_-{\ones{n}}
}
\end{equation}

Again, let $\mu$ denote the image of $\mu(1)$ in $M$. The above
diagram then implies
\begin{equation}
  \langle \mu,\alpha\rangle=1, \quad\forall \alpha\in\cA  \label{eq:Aplane}
\end{equation}

The \GLSM s of interest are thus those for which 
$\cA$ and $\cB$ both lie in
``primitive'' hyperplanes in the sense that we have lattice vectors
$\mu$ and $\nu$ defining the hyperplanes (\ref{eq:Bplane}) and
(\ref{eq:Aplane}).\footnote{This condition also shows that the points
  $\alpha\in\cA$ are primitive generators of the rays in the toric
  fan. This means we avoid having to use any {\em stacky\/}
  construction of the form described in \cite{BCS:toric}.}

\subsection{The Low-Energy Limit}

A collection of data satisfying this condition will determine 
a family of \GLSM s, whose extreme IR limits will be a family of
$\cN{=}(2,2)$ \SCFTS\ of central charge given by \cite{AG:gmi}
\begin{equation}
  \ff13 c = d-2\langle\mu,\nu\rangle.
\end{equation}
(This is also the dimension of the associated \CY\ category.)
The deformation theory is simple and well understood.  The
deformation space splits locally as a product of two complex special
K\"ahler spaces associated to deforming the action by the top
components of chiral (resp. twisted chiral) multiplets of charge
$(1,1)$ (resp. $(1,-1)$) under the left- and right-moving $R$-symmetry.
Deformations are not obstructed: all supersymmetric marginal
deformations are truly marginal.  

With our conventions, the deformations associated to chiral multiplets
will be holomorphically parametrized by a choice of the coefficients
$b_\beta\in\C^m$ determining the superpotential.  Since deformations
of the kinetic terms are irrelevant in the IR theory, changes in these
that can be undone by field redefinitions consistent with the
R-symmetry are redundant.  
We will here restrict attention to diagonal field
redefinitions contained in $(\C^*)^n$.  These act on $\C^m$ through an
action determined by $P$, whose rank shows that the space of orbits
has dimension $m-d$. As is standard, we define a compact model for
this space of deformations as the toric variety associated to the
{\em secondary fan\/} of $\cB$. We denote this compact space $\cM_B$.

Deformations associated to
twisted chiral multiplets will be parametrized by the \FI\ coefficients and
$\theta$ angles for the gauge group, both of which live in
$\fkg^*_{\R}$. The combination $\tau =
i\rho + \frac{\theta}{2\pi}\in\fkg^*$ will give local holomorphic coordinates.
This space $\cM_A$ will thus have dimension $r = \rank G = n-d$.
Again we define $\cM_A$ to be the toric variety associated to
the secondary fan of $\cA$. Locally $q = \exp(2\pi i\tau)$ are good coordinates.

Note we make no claim that $\cM_B$ is the {\em moduli space\/} of
chiral deformations, nor $\cM_A$ the moduli space of 
twisted chiral deformations. The dimensions will typically be
incorrect.  For $\cM_B$ this
may be because of non-polynomial deformations, non-toric field
redefinitions, or simply because we
choose to not include a point in $\cB$ that could have been used as a
monomial in $W$. Similar statements apply to $\cM_{A}$ and $\cA$.

For special values of the parameters of either type the low-energy
theory will be singular.   Because of the factorization properties the
{\em singular locus\/} will locally be the product of two complex
subvarieties of $\cM_B$ and $\cM_A$ (although at intersections of
these the singularity type changes in interesting ways this will not
concern us here).  We will partially characterize these subvarieties
below.   In particular, we wish to formulate conditions on the
combinatorial data determining the family sufficient to ensure that
the generic member is not singular.

We can predict the singular locus by considering the space of
gauge-inequivalent supersymmetric field configurations.
These are classically determined by
constant expectation values for the scalars $x_\alpha$ in the chiral
multiplets and a field $\sigma\in \fkg$ in the twisted chiral
multiplets for which the potential energy density given by
\begin{equation}
\begin{split}
U(x,\sigma) &= |D|^2 + \sum_\alpha |F_\alpha|^2 + \sigma^\dagger M(x)\sigma\\
D &= \sum_\alpha Q_\alpha |x_\alpha|^2 - \rho\in \frak{g}_{\R}^*\\
F_\alpha &= \frac{\p W}{\p x_\alpha}\\
M &= \sum_{\alpha} {}^tQ_\alpha Q_\alpha |x_\alpha|^2\in
      \frak{g}_{\R}^{\otimes 2}\label{eq:ubos}
\end{split}
\end{equation}
vanishes.  Solutions related by the action of the compact real form of
the gauge group $G$ via $q$ are identified (the action on $\sigma$
is trivial).

The solution for $x$ here is familiar.
Values of $\rho$ for which $M$ has a kernel in the space of solutions
to (\ref{eq:ubos}) lie in cones and form the
faces of a fan in $\frak{g}_{\R}^* = {\mathbb R}^r$, the secondary fan
of $\cA$.
Large cones in this fan are associated
to a choice of triangulation of $\cA$  which, in turn, determines a fan
$\Sigma$. This also gives an associated irrelevant ideal $B$ in the
homogeneous coordinate ring $S=\C[x_1,\ldots,x_n]$.
We then get the usual toric variety
\begin{equation}
  Z_\Sigma = \frac{\Spec S - V(B)}{G}.
\end{equation}
The hyperplane condition (\ref{eq:Aplane}) implies that $Z_\Sigma$ is
not compact and its canonical divisor $K_{Z_\Sigma}$ is zero. However,
at this stage, this divisor need not be Cartier and thus $Z_\Sigma$
need not be Gorenstein. We do not know a physical reason why one
should impose the condition of being Gorenstein on $Z_\Sigma$.

For values of $\rho$ deep inside one of these maximal cones, the gauge
symmetry is broken at high energies and a semiclassical description is
valid.  From (\ref{eq:ubos}) the space of vacua is the critical set of
$W$.  In general there are massless fields in the low-energy theory
interacting via superpotential interactions and the vacuum structure
is that of a ``bad'' hybrid theory.\footnote{That is, a
  ``pseudo-hybrid'' in the sense of \cite{me:hybridm}.} Suppose this
critical set has a non-compact component extending to arbitrarily
large expectation values for the fields.  Near these solutions
semiclassical considerations are valid and the noncompact space
supports a continuum of states extending down to zero energy. This
leads to {\em singular\/} low-energy behavior \cite{W:phase}.

Similarly values of $\rho$ lying in the
faces of the secondary fan can also lead to classically singular IR physics.
In the \GLSM\ this singularity is evident in that the space of
supersymmetric field configurations 
includes configurations for which $M(x)$ has a kernel.  This means a
continuous subgroup of $G$ is not broken and the associated $\sigma$
field is free.  This introduces a noncompact component into the space
of vacua, and again far out along this component semiclassical
considerations are a good approximation and we expect the continuous
spectrum to extend down to zero energy and lead to singular IR behavior.  

These classical considerations are modified by quantum corrections,
which we discuss in section \ref{ss:GKZ-A}, but in the regions of
large $\sigma$ expectation values these are very well controlled.  In
principle these considerations cannot exclude the possibility that
there are intrinsically strongly coupled singularities that are not
determined by semiclassical considerations.  This can be demonstrated
by using the existence of two twisted topological versions of these
models, the {\bf A} and {\bf B} models of \cite{W:AB}. In the
topological models semiclassical calculations are exact and lead to
the conclusions described in detail in the following two subsections.

We thus arrive at our criterion for a theory being nonsingular. The
space of classical vacua given by (\ref{eq:ubos}) as corrected below
must be compact.  A generic member of the family will be nonsingular
as long as the singular locus is of codimension at least one in
$\cM_A$ and $\cM_B$.

\subsection{Geometry} \label{ss:geom}

A natural question to ask is when $X$ is a \CY\ variety and so our
theory has an interpretation as a nonlinear $\sigma$-model at large
radius. This turns out to be a surprisingly awkward question.

The strongest constraint one can impose is that there is a ``nef
partition''. Let $s=\langle\mu,\nu\rangle$. In \cite{MR2405763} it was
shown that this is equivalent (assuming we are in the reflexive case)
to both $\cA$ and $\cB$ admitting a ``special $(s-1)$-simplex''. That
is, there is an $(s-1)$-simplex with its $s$ vertices in $\cA$ such
that each facet of $\Conv\cA$ contains $s-1$ vertices of the simplex;
and similarly for $\cB$. There is then a triangulation, $\Sigma$, of
$\cA$ such that each simplex in the triangulation has the special
$(s-1)$-simplex as a face. Then $X_\Sigma$ will be a complete
intersection of $s$ hypersurfaces and a \CY\ variety. Furthermore, the
mirror of $X_\Sigma$ will also be a complete intersection \CY.

Dropping this nef partition condition, we could still have $\cA$
admits a special $(s-1)$-simplex but $\cB$ doesn't. Then $\cA$ could
still have a triangulation leading to a \CY\ complete intersection but
its mirror does not. An example of this is the Z-manifold
\cite{AG:gmi}.

Going further, it might be that neither $\cA$ nor $\cB$ admit a
special $(s-1)$-simplex. Then there is no ``obvious'' geometric
phase but there are still ways that $X$ could really be geometric. For
example
\begin{enumerate}
\item $Z$ contains more than one component in its compact part, like
  an exoflop, but $\Crit(W)$ misses all but one of these
  components. Thus $X_\Sigma$ looks like a complete intersection again
  even though the triangulation $\Sigma$ is not associated to a
  $(s-1)$-simplex.
\item There could be a large radius limit point in $\cM_A$ that is not
  a phase limit point. Such cases, associated to flops, were
  discussed in \cite{GMV:Ht,AdAs:masscat}. More general cases are
  presumably possible. In particular, large radius limits can hide on
  the singular locus
  deep inside the moduli space, away from phase limits.
\end{enumerate}

Note that we can always {\em force\/} a geometric interpretation by
including $\nu$ as an extra ray in the fan $\Sigma$. If $\nu$ is
interior to $\cA$ this puts a divisor in the new $Z_\Sigma$ and our
new quasi-$X_\Sigma$ will be a variety of dimension $d-2$. If $s>1$ it
will not be \CY. Such a construction is common in the literature
(starting with the cubic sevenfold of \cite{Drk:Z} as the mirror of
the Z-manifold) but we want to claim it is unnatural
in the context of the \GLSM.\footnote{Not that it is without
  mathematical virtue. For example, the \CY\ category of $X$ appears
  as a factor in the semiorthogonal decomposition of the derived
  category of the $(d-2)$-fold. See section \ref{sec:exm}.} Indeed,
the extremal transitions between geometric and non-geometric models in
section \ref{sec:exm} will show no signs of this $(d-2)$-fold.

\subsection{The GKZ Determinant for $\cB$} \label{ss:GKZ-B}

\subsubsection{The generic $\C^*$ orbit}

Let us consider first the family of complex structures parametrized by
$\cM_B$.  The singular locus in $\cM_B$ is not subject to quantum
corrections and can be determined directly from the space of classical
vacua.  For generic values of the complexified FI parameters,
supersymmetric field configurations will be given by an assignment of
constant expectation values to the scalars in the chiral multiplets
$x$, identified up to the action of $G$.

Within $Z_\Sigma$ we have the classical vacuum space $X_\Sigma$ given
by the critical locus of $W$. When 
$X_\Sigma$ is non-compact the model is singular.
The choice of superpotential for which the theory is singular
should not depend at all on the choice of twisted chiral parameters $\tau$,
so we
should not need to choose a fan $\Sigma$.

The torus $T_N$ acts on $Z_\Sigma$ but does not generally fix
$X_\Sigma$. However, because of the $R$-symmetry, we know that the
one-parameter subgroup $\lambda^\nu:\C^*\to T_N$ given by $\nu\in N$
will preserve $X_\Sigma$. We can use the cone-orbit correspondence
\cite{CLS:ToricVar} to see how $\lambda^\nu(\C^*)$ acts on $Z_\Sigma$.
In particular, suppose $x\in X_\Sigma$ is a point whose homogeneous
coordinates are all nonzero. Then the cone-orbit correspondence tells
us that 
\begin{equation}
  \lim_{t\to 0} \lambda^\nu(t)x,
\end{equation}
lies within $X_\Sigma$ if and only if $\nu$ lies in the fan
$\Sigma$. Similarly
\begin{equation}
  \lim_{t\to\infty} \lambda^\nu(t)x,
\end{equation}
lies within $X_\Sigma$ if and only if $-\nu$ lies in the fan
$\Sigma$. But the hyperplane condition for $\cA$ means that $\nu$
and $-\nu$ cannot both be in $\Sigma$ and so the orbit
$\lambda^\nu(\C^*)$ cannot be compact.

Thus we have
\begin{prop}
If $X_\Sigma$ contains a point with all homogeneous coordinates
nonzero then $X_\Sigma$ cannot be compact and the associated \GLSM\ is
singular.
\end{prop}

GKZ \cite{GKZ:book} have defined a discriminant which exactly
determines when this happens. Let $z_1,\ldots,z_d$ denote affine
coordinates on $Z_\Sigma$ which may be constructed as Laurent monomials in the
homogeneous coordinates using the matrix $A$. Then $W$ may be written as
a Laurent polynomial in $z_j$'s. Let $\nabla_0\in\C^{\cB}$ denote the
set of all $W$ for which there exists a point with nonzero $z_j$'s
such that
\begin{equation}
 \frac{\partial W}{\partial z_j}=0,\quad\forall j,
\end{equation}
and let $\nabla_{\cB}$ denote the closure. (The homogeneity of $W$,
forced by the $R$-symmetry, will imply $W=0$ automatically.) Assuming
$\nabla_{\cB}\subset\C^{\cB}$ is codimension one, this defines a
polynomial $\Delta_{\cB}$ in the $b_\beta$'s. This is the
``$\cB$-discriminant'' and tells us when the \GLSM\ becomes singular
because of the above $\C^*$-orbit.

Using the Horn uniformization of \cite{GKZ:book} and, as we will
discuss more carefully in section \ref{sss:Aprim},
we can use the fact that $\cM_B$ is (possibly redundantly)
parametrized by $b_\beta$ up to rescalings of the $x_\alpha$ to state
this condition more directly in terms of the coefficients.  The \GLSM\
will be singular for any values of $b_\beta$ equivalent up to
rescaling to the solution of 
\begin{equation}\label{eq:dinbs}
\sum_{\beta\in\cB}\la \beta,\alpha\ra b_\beta =
0\quad\forall\alpha\in\cA\ ,
\end{equation}

\subsubsection{Special Orbits}  \label{sss:faces}

But this is not the only way to find singularities. We should also
consider setting some of the $x_\alpha$'s equal to zero, thus killing
some of the terms in the superpotential. Consider a subset of
coordinates we wish to set to zero. This corresponds to a subset of
the pointset $\cA$. If the rays through these points do not span a
cone, $\tau$, then we cannot simultaneously set these coordinates to
zero.

Assuming $\tau$ exists, we restrict to the subvariety corresponding to
the zeroes of these coordinates by quotienting $N_\R$ by the span of
the rays through these points. The sub-toric variety is then determined
by the fan given by Star$(\tau)$ (see proposition 3.2 of
\cite{CLS:ToricVar}).

If the fan Star$(\tau)$ is complete we have a compact toric subvariety
and so we need not concern ourselves with the possibility of
decompactifying the classical vacuum $\Crit(W)$. In this case, the
$\C^*$ action given by the R-symmetry can have nontrivial orbits which
are compact.

The case that we still need to worry about is when $\tau$ lives in a
{\em face\/} of the convex hull of the pointset $\cA$, $\Conv{\cA}$. It
is then that Star$(\tau)$ is not complete.

So let $\Upsilon$ denote a face of $\Conv{\cA}$ and let
$\Upsilon^\perp$ denote the points in $M_\R = M \otimes_\Z\R$
perpendicular to this. Then, upon setting the homogeneous coordinates
in the face $\Upsilon$ to zero, the only terms in the superpotential
which remain nonzero correspond to $\cB\cap\Upsilon^\perp$. Subsets of
$\cA$ which do not correspond to faces of $\Conv{\cA}$ do not yield
any other possibilities.

Let $\Cone(\Conv\cA)^\vee$ denote the cone in $M_\R$ dual to the
cone over $\Conv\cA$. We have just shown that an interesting
superpotential to consider is one where the nonzero monomials lie in a
face of $\Conv\cB$ which lies in the boundary of $\Conv^\vee\cA$. Let
$\Gamma$ denote the such a face of $\Conv\cB$. Then noncompact orbits
of the R-symmetry $\C^*$ action can occur only if the discriminant
$\Delta_{\cB\cap\Gamma}$ vanishes.

This is getting very close to reproducing the GKZ
``$\cB$-determinant''. This is defined as \cite{GKZ:book} as resultant
polynomial, $E_{\cB}$, associated to the simultaneous vanishing of
\begin{equation}
x_1\frac{\partial W}{\partial x_1}, \;x_2\frac{\partial W}{\partial
  x_2},\;\ldots,\;x_n\frac{\partial W}{\partial x_n}. \label{eq:Ed}
\end{equation}
It is shown that \cite{GKZ:book} 
\begin{equation}
  E_{\cB} = \prod_{\Gamma} \Delta_{\cB\cap\Gamma}^{u(\Gamma)},  \label{eq:GKZd}
\end{equation}
where the product is over the faces of $\Conv{\cB}$. To be precise,
faces are included by the following rules starting from faces of
dimension zero and working upwards:
\begin{enumerate}
\item All vertices are included.
\item Higher dimensional faces are included only when they add points
  obeying nontrivial affine relations.
\end{enumerate}
For example, if $\Conv\cB$ were a simplex and only had points at the
vertices, only the vertices would be included in the product
(\ref{eq:GKZd}). Higher dimensional faces appear in the product if
they are non-simplicial or have internal points. The multiplicity
$u(\Gamma)$ is a positive integer which, while potentially very
interesting, will not concern us in this paper. We show some simple
examples in figure \ref{fig:Deg}.

\begin{figure}
\begin{center}
\begin{tikzpicture}[scale=1.0]
\filldraw (0,0) circle (0.05) node[anchor=east] {$a$};
\filldraw (1,0) circle (0.05) node[anchor=west] {$c$};
\filldraw (0.5,0) circle (0.05) node[anchor=south] {$b$};
\draw (0.2,-1.0) node {$E_{\cB}=ac(b^2-4ac)$};
\end{tikzpicture}\hspace{9mm}
\begin{tikzpicture}[scale=1.0]
\filldraw (0,0) circle (0.05) node[anchor=west] {$c$};
\filldraw (1,0) circle (0.05) node[anchor=west] {$b$};
\filldraw (0.5,0.886) circle (0.05) node[anchor=west] {$a$};
\draw (0.5,-1.0) node {$E_{\cB}=abc$};
\end{tikzpicture}\hspace{9mm}
\begin{tikzpicture}[scale=1.0]
\filldraw (0,0) circle (0.05) node[anchor=west] {$c$};
\filldraw (1,0) circle (0.05) node[anchor=west] {$b$};
\filldraw (0.5,0.886) circle (0.05) node[anchor=west] {$a$};
\filldraw (0.5,0.3) circle (0.05) node[anchor=west] {$d$};
\draw (0.5,-1.0) node {$E_{\cB}=a^2b^2c^2(d^3+27abc)$};
\end{tikzpicture}
\end{center} \caption{Some examples of $E_{\cB}$ for a given $\cB$.} \label{fig:Deg}
\end{figure}

\begin{definition}
The pointsets $\cA$ and $\cB$ are {\em reflexive\/} if
\begin{equation}
  \Cone(\Conv\cA)^\vee = \Cone(\Conv\cB).
\end{equation}
\end{definition}
In this case there is a correspondence between faces of $\Conv\cA$ and
dual faces of $\Conv\cB$. So the product over the faces of $\Conv\cB$
in (\ref{eq:GKZd}) corresponds to a product over faces of $\Conv\cA$,
i.e., setting various combinations of $x_\alpha$'s to zero. Thus, if
$E_{\cB}$ is nonzero then the $\C^*$ orbits acting within $\Crit(W)$
associated to the R-symmetry will be compact.

We now invoke a key result from GKZ. Namely we use proposition 1.1 of
chapter 10 of \cite{GKZ:book}, which can be stated as
\begin{prop}
  If $M_\R$ is generated\,\footnote{GKZ assume the stronger condition
    that $\cB$ affinely generates $M\cap H_\nu$. This is the same as
    our assumption up to a torsion group but that is not important for
    the statement that the determinant is generically nonzero.} by
  $\cB$ then the variety determined by simultaneous vanishing of
  (\ref{eq:Ed}) is codimension one and corresponds to the determinant
  function $E_\cB$. This determinant is generically nonzero.
\end{prop}
The reflexive condition forces $\cA$ to generate $M_\R$ and so we have
\begin{prop}
  If the pointsets $\cA$ and $\cB$ are reflexive then the $\C^*$
  orbits acting within $\Crit(W)$ associated to the R-symmetry will be
  compact for generic values of the coefficients $b_\beta$ in the
  superpotential.
\end{prop}

\subsubsection{A sufficient condition for compactness}

Any time we have a nontrivial R-symmetry $\C^*$ orbit, the theory is
noncompact and thus singular.  We would now like to prove that the
{\em only\/} way that the \GLSM\ can become singular is when we have
such orbits. That is, if $\Crit(W)$ is fixed by the R-symmetry, we need
to show it is compact.

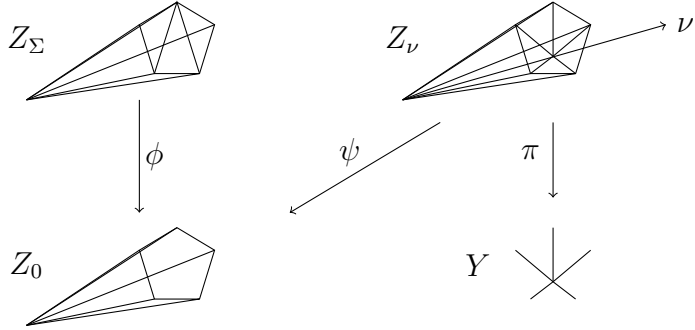
\begin{figure}
\begin{center}
\begin{tikzpicture}[scale=1.0]
\coordinate (a0) at (0.5,0);
\draw (a0) -- (2,1);
\draw (a0) -- (3,1);
\draw (a0) -- (2.5,1.3);
\draw (a0) -- (2.2,0.35);
\draw (a0) -- (2.8,0.35) -- (2.2,0.35) -- (2,1) -- (2.5,1.3) -- (3,1)
           -- (2.8,0.35);
\draw(0.5,0.8) node {$Z_0$};
\begin{scope}[yshift=3cm]
\draw[->] (2,0) -- (2,-1.5);
\draw (2.2,-0.75) node {$\phi$};
\coordinate (a0) at (0.5,0);
\draw (a0) -- (2,1);
\draw (a0) -- (3,1);
\draw (a0) -- (2.5,1.3);
\draw (a0) -- (2.2,0.35);
\draw (a0) -- (2.8,0.35) -- (2.2,0.35) -- (2,1) -- (2.5,1.3) -- (3,1)
           -- (2.8,0.35);
\draw (2.2,0.35) -- (2.5,1.3) -- (2.8,0.35);
\end{scope}
\draw(0.5,3.8) node {$Z_\Sigma$};
\begin{scope}[xshift=5cm,yshift=3cm]
\coordinate (a0) at (0.5,0);
\draw (a0) -- (2,1);
\draw (a0) -- (3,1);
\draw (a0) -- (2.5,1.3);
\draw (a0) -- (2.2,0.35);
\draw (a0) -- (2.8,0.35) -- (2.2,0.35) -- (2,1) -- (2.5,1.3) -- (3,1)
           -- (2.8,0.35);
\draw[->] (a0) -- (4,1) node[anchor=west] {$\nu$};
\coordinate (c0) at (2.5,0.58);
\draw (2,1) -- (c0);
\draw (3,1) -- (c0);
\draw (2.5,1.3) -- (c0);
\draw (2.2,0.35) -- (c0);
\draw (2.8,0.35) -- (c0);
\end{scope}
\draw(5.5,3.8) node {$Z_\nu$};
\draw[->] (6,2.7) -- (4,1.5);
\draw(4.8,2.3) node {$\psi$};
\begin{scope}[xshift=5cm]
\coordinate (c0) at (2.5,0.58);
\draw (2,1) -- (c0);
\draw (3,1) -- (c0);
\draw (2.5,1.3) -- (c0);
\draw (2.2,0.35) -- (c0);
\draw (2.8,0.35) -- (c0);
\end{scope}
\draw(6.5,0.8) node {$Y$};
\draw[->] (7.5,2.7) -- (7.5,1.7);
\draw(7.2,2.3) node {$\pi$};
\end{tikzpicture}
\end{center} \caption{$\Crit(W)$ is compact if it is fixed by the
  R-action.} \label{fig:Z}
\end{figure}

To do this we consider various maps as shown in figure
\ref{fig:Z}. Let $Z_0$ be the toric variety associated to the cone
over $\Conv(\cA)$. We therefore have a toric map $\phi:Z_\Sigma\to
Z_0$, which contracts various toric subspaces. 

It is clear that reflexive condition forces $\nu$ to lie in the proper
interior the $\Cone(\Conv(\cA))$. This means we can consider another
toric variety $Z_\nu$ obtained from a star subdivision along the ray
passing through $\nu$ as shown in figure \ref{fig:Z}. This gives a
toric map $\psi:Z_\nu\to Z_0$ contracting the divisor associated to
$\nu$.

Consider the $\C^*$-action associated to $\nu\in N$. This
extends to a well-behaved $\C$-action on $Z_\nu$. Quotient $Z_\nu$ by
this $\C$ by projecting the lattice $N$ onto the quotient lattice
$N/\nu$. This projects the fan to produce a new toric variety $Y$. 
The resulting map 
\begin{equation}
\pi:Z_\nu\to Y,
\end{equation}
is again toric. Because $\nu$ was in the interior of
$\Cone(\Conv(\cA))$, the resulting fan is complete and thus $Y$ is
{\em compact}.  Now $\Crit(W)$ is an algebraic subset and thus
closed. Therefore its image in the quotient $Y$ is
compact. Furthermore, the map $\phi$ is {\em proper\/} from theorem
3.4.11 of \cite{CLS:ToricVar}. This means the inverse image of a
compact set in $Z_0$ is compact in $Z_\Sigma$. It follows that if
$\Crit(W)$ has only compact orbits under the R-symmetry $\C^*$-action,
by chasing the maps in figure \ref{fig:Z}, it must be compact in
$Z_\Sigma$.

We have therefore proven:
\begin{prop}  \label{p:main-B}
For the fields so far considered, for a generic choice of
coefficients, $b_\beta$, the \GLSM\ is nonsingular if the pointsets
$\cA$ and $\cB$ are reflexive.
\end{prop}

The singular models are contained in the locus $E_\cB=0$. As we will
see, however, this condition is too strong in general, and there are
typically many smooth models for which $E_\cB=0$. Furthermore $\cA$
and $\cB$ do not need to be reflexive to give a generically smooth model.

Note that the previous motivation for reflexivity in
\cite{Bat:m,Boris:m} was that $Z_\Sigma$ should be a vector bundle
over a Gorenstein variety. $X_\Sigma$ (assuming it is geometric) is then
viewed as a complete intersection in the Gorenstein variety.

\subsection{The Determinant for $\cA$} \label{ss:GKZ-A}

\subsubsection{The Primary Component}  \label{sss:Aprim}

The considerations of the previous section determined conditions under
which the family of \GLSM s includes sufficiently general
superpotentials that the generic member will not be singular, assuming
that $Z_\Sigma$ is sufficiently generic.  A key observation was that the
singular locus in $\cM_B$ is independent of the twisted chiral
parameters.  In this section our goal is to understand the singular
locus in $\cM_A$ and the combinatorial conditions under which the
generic element of the family avoids this.   
As mentioned above, a tree level analysis finds singular IR limits and
a non-compact space of vacua for values of $\rho$ lying in the faces
of the secondary fan.  

These classical expectations are corrected by quantum effects as found
in \cite{W:phase,MP:inst}.  These introduce a dependence on the periodic
$\theta$ angles for the gauge group.  The IR physics in fact depends
holomorphically on the combination $q_j = e^{2\pi i\tau_j}$.
The scalar fields $\sigma$ are contained in twisted chiral multiplets
and their interactions controlled by a holomorphic twisted
superpotential whose critical points are the vacua.  The classical
superpotential 
\begin{equation}
\widetilde W_0 = \sum_{j=1}^r \Sigma_j\log q_j 
\end{equation}
leads to the cone structure described above, and we can now see how
quantum effects modify this.

The noncompact space of vacua responsible for the singular low-energy
physics lie at large values of $\sigma$. If $\sigma$ is nonzero then
(\ref{eq:ubos}) can lead to mass terms for the chiral fields.  Let us
suppose, for the time being, that $\sigma$ takes a value that is
sufficiently generic so that {\em all\/} the $x_\alpha$'s become
massive. We deal with the more general case in section
\ref{sss:Afaces}. Integrate out these massive fields to
produce a correction to the twisted superpotential
\begin{equation}
\widetilde W_1 = -\sum_{j=1}^r\Sigma_j\left[\sum_{\alpha=1}^n
  Q_\alpha^j \log\left(\sum_{k=1}^r Q_\alpha^k \Sigma_k\right)\right]\
.
\end{equation}
Critical points of the twisted superpotential $\widetilde W_0 +
\widetilde W_1$ satisfy 
\begin{equation}
q_j = \prod_\alpha \left(\sum_k
  Q_\alpha^k\sigma_k\right)^{Q_\alpha^j}\ . \label{eq:primd}
\end{equation}
These are a set of $r$ homogeneous equations for the $r$ expectation
values $\sigma_j$.  For generic $q_j$ there will be no solutions,
while for values of $q_j$ for which a solution exists there will be a
noncompact {\em flat direction} for $\sigma$.  We can consider these
as parametric equations determining values of $q_j$ for which the
model is singular. This exactly reproduces the ``Horn uniformization''
of equation (3.3) in for the $A$-discriminant in \cite{GKZ:book}. If
this singular set is codimension one (as it is in most examples) we
call it the ``primary component'' of the singular locus.

To compare this to the expressions from the previous section it is 
helpful to express the parameters $q_j$ in terms of a redundant
set $a_\alpha$ as
\begin{equation}\label{eq:gkza}
q_j = \prod_\alpha a_\alpha^{Q_\alpha^j}\ ,
\end{equation}
expressing them as affine coordinates on a quotient of $\C^n$
by $(\C^*)^d$.    
In terms of the $a_\alpha$ we see that the solutions to 
(\ref{eq:primd}) are precisely the orbits under this action containing
a solution to 
\begin{equation}
a_\alpha = \sum_k Q_\alpha^b\sigma_k\ .
\end{equation}
The sequence (\ref{eq:kerQ}) shows that this is equivalent to 
\begin{equation}
\sum_{\alpha\in\cA}\la \beta,\alpha\ra a_\alpha =
0\quad\forall\beta\in\cB\ .
\end{equation}

The locus parametrized by (\ref{eq:primd}) has large\,-$|\rho|$ limits
which asymptote, up to a finite shift, to some of the faces of the
secondary fan.  This is expected in the sense that in these limits the
quantum corrections are suppressed.  In these limits the solution for
$\sigma$ lies in the direction of the symmetry that is unbroken in the
classical calculation.  When $\sigma$ points in these special
directions, (\ref{eq:ubos}) leaves a subset of the chiral fields
$x_\alpha : Q_\alpha^j\sigma_j = 0$ 
massless.  In the
interior of $\rho$-space, the true singular locus interpolates smoothly
between these various limits. At a generic point on it the $\sigma$
vacua are in a generic direction in $\sigma$-space
leading to nonzero masses for all of the chiral fields.  These
singularities are insensitive to the superpotential, since the chiral
fields are massive.

\subsubsection{Other Components} \label{sss:Afaces}

So far we have considered vacua where either $x$'s are
interesting and the $\sigma$'s are all zero, i.e., Higgs branches; or
vacua where the $\sigma$'s are nonzero and the $x$'s are all massive,
i.e., Coulomb branches. Now consider the mixed cases where the $\sigma$
fields can be nonzero but some $x$'s remain massless. This will lead
us to looking at faces of $\Conv\cA$ as we now explain.

To understand these other components we first ignore the
superpotential and seek loci in which there are such mixed branches at
$W=0$. 

Choose a subgroup $\widehat G_h\subset\widehat G$ and let $\widehat
G_c$ be the quotient $\widehat G/\widehat G_h$, with $\pi_h$ the
quotient map. Let $e_\alpha$ be a basis for $\Z^{\cA}$. Then define
\begin{equation}
\begin{split}
\cA_h &= \{ \alpha\in\cA: \pi_h Qe_\alpha=0\}\\
\cA_c &= \cA - \cA_h.
\end{split}
\end{equation}
If we put $\fkg_c=\Hom(\widehat G_c,\C)$, then we have partitioned
$\cA$ into $\cA_h$, which are fields neutral under $\fkg_c$; and
$\cA_c$, which are charged.  
\begin{prop} There is a commutative diagram
\begin{equation}
\xymatrix{
&0&0&0\\
0\ar[r]&M_c\ar[r]\ar[u]&\Z^{\cA_c}\ar[r]^-{Q_c}\ar[u]&
  \widehat G_c\ar[u]\ar[r]&0\\
0\ar[r]&M\ar[r]^{A^t}\ar[u]&\Z^{\cA}\ar[r]^Q\ar[u]&
  \widehat G\ar[u]_{\pi_h}\ar[r]&0\\
0\ar[r]&M_h\ar[r]\ar[u]&\Z^{\cA_h}\ar[u]\ar[r]^-{Q_h}&
   \widehat G_h\ar[u]\ar[r]&0\\
&0\ar[u]&0\ar[u]&0\ar[u]\\
}
\end{equation}
with all rows and columns exact.
Here $Q_c$ is defined as $\pi_hQ$ restricted to $\Z^{\cA_c}$, and
$Q_h$ exists uniquely to make the diagram commute. $M_c$ and $M_h$ are
defined as the kernels of $Q_c$ and $Q_h$ respectively.
\end{prop}
To prove this proposition note that the first two columns consist of
free groups; then use projectivity to prove the existence of the
maps. Then use the Snake Lemma to prove the first column is exact.\QED

Now assume that $\sigma$ acquires a large generic value in $\fkg_c$,
which generates masses for all the chiral fields in $\cA_c$.
Integrating out these fields associated to $\cA_c$ will produce an
effective twisted superpotential for $\sigma$ and a correction to the
$D$-terms in $\fkg_c$.  The $D$-terms for $\fkg_h$ will not be
corrected. The resulting model decomposes into a ``product'' of a twisted
Landau-Ginzburg theory for $\sigma\in\fkg_c$ and a reduced \GLSM\ involving
the chiral fields associated to points in $\cA_h$ and the reduced
gauge group $G_h$.

So we have a mixed Higgs--Coulomb branch where the points in $\cA_h$
are associated to the Higgs data and the points in $\cA_c$ are
associated to the Coulomb data. The primary component in the previous
section corresponds to the case $\widehat G_h=\widehat G$; i.e.,
$\cA_h=\emptyset$ and $\cA_c=\cA$.

The Landau-Ginzburg model for the Coulomb branch will have
supersymmetric vacua, hence flat directions leading to singular IR
physics, when the FI parameters in $\fkg_c^*$ satisfy
\begin{equation}
q_{j} = \prod_{\alpha\in\cA_c} \left(\sum_{k}
  Q_{c,\alpha}^k\sigma_k\right)^{Q_{c,\alpha}^{j}}\ .  \label{eq:HR1}
\end{equation}
We have an exact sequence
\begin{equation}
\xymatrix@1{
0 \ar[r]&\fkg^*_h\ar[r]&\fkg^*\ar[r]&\fkg^*_c\ar[r]&0},
\end{equation}
so we have a singular theory if the K\"ahler parameters of the total
theory lie in the inverse image of the K\"ahler parameters of the
Coulomb theory satisfying (\ref{eq:HR1}).

Now we turn our attention to the \GLSM\ associated to the Higgs
data. The points $\cA_h$ should be viewed as lying in the quotient
lattice $N_h = N/N_c$. The first thing to observe is that this
generally destroys the \CY\ condition --- the points will not lie in a
hyperplane, and so we should take care to see where the IR flow lands
us. Let us build the secondary fan $F_h$ for the Higgs data
$\cA_h\subset N_h$ in the usual way (see, for example, section 15.2 of
\cite{CLS:ToricVar}). There is no reason the suppose this fan is
complete. The case of an incomplete fan for non-\CY\ data is familiar
from \cite{W:phase,MP:inst}. Here the IR flow pushes us beyond the
exterior walls of the fan to where supersymmetry is na\"\i vely
broken. More careful analysis shows that supersymmetry is unbroken,
but there are nonzero expectation values for some of the $\sigma$
fields living in $\fkg_h$. So, in this case, our model is just
reproducing a different Coulomb--Higgs split where $\cA_c$ is larger.

To avoid reproducing multiple copies of the same components of the
discriminant we should restrict attention to case where the secondary
fan $F_h$ is {\em complete}. Now the region covered by the secondary fan
is spanned by positive combinations of columns of $Q_h$.  Let
$\mathbf{x}_\alpha$ be the position vector of the point $\alpha\in\cA_h$ in
$N_h$. So the fan cannot be complete if there is any
nontrivial relationship
\begin{equation}
  \sum_{\alpha\in\cA_h} Q_\alpha\mathbf{x}_\alpha=0,
\end{equation}
where all the $Q_\alpha$'s are $\geq0$.  This imposes two conditions:
\begin{enumerate}
\item No point in $\cA_h\subset N_h$ should lie at the origin.
  $\cA_c$ spans the subspace $N_{c,\R}\subset N_{\R}$ and so we
  require $\cA_h\cap N_{c,\R}=\emptyset$. That is,
  $\cA_c=N_{c,\R}\cap\cA$.
\item The subspace $N_{c,\R}$ must meet $\Conv\cA$ along a {\em face}.
\end{enumerate}

We have therefore almost recovered the GKZ prescription of considering
all the faces of the convex hull of $\cA$ similar to
(\ref{eq:GKZd}). There is a notable difference, however, when we ask
exactly which faces are included. We include faces given by $\cA_c$
when, and only when, there is a $\fkg_c$ such that $\cA_c$ is the set
of fields with nonzero charges. Working upwards from lower dimensions,
these are precisely the faces with a new nontrivial affine relation. We
therefore reproduce the GKZ rules of section \ref{sss:faces}, {\em
  except we have no vertices in the product.}  We return to this issue
shortly.

Anyway, we can use the results of GKZ again to prove that the model is
nonsingular for a generic choice of K\"ahler data so long as $\cA$
generates $N_\R$. So we only need modify proposition \ref{p:main-B}
slightly to obtain:

\begin{theorem} \label{th:main} For a generic choice of coefficients,
  $b_\beta$, and values of $i\rho_j+\frac{\theta_j}{2\pi}$, the \GLSM\ is
  nonsingular if the pointsets $\cA$ and $\cB$ are reflexive.
\end{theorem}

\subsection{Mirror Symmetry} \label{ss:mir}

There is an obvious symmetry in the above construction that
is, of course, mirror symmetry:
\begin{equation}
\begin{split}
N &\leftrightarrow M\\
\nu &\leftrightarrow \mu\\
\cA &\leftrightarrow \cB\\
a_\alpha&\leftrightarrow b_\beta\\
\cM_A&\leftrightarrow \cM_B
\end{split} \label{eq:mir1}
\end{equation}
Let $Y$ refer to the mirror of $X$ when the above mirror switch is
applied. 

In particular, the last of the identifications in (\ref{eq:mir1}) is
an very interesting pointwise map between our models of the K\"ahler
moduli space of $X$ and then complex structure moduli space of $Y$
(and vice versa). It was first proposed in \cite{MP:inst} and is a
natural extension of the ``monomial-divisor mirror map'' of
\cite{AGM:mdmm}. Asymptotically, in a large radius limit of $X$ and
near a point of maximally unipotent monodromy, it becomes the usual
mirror map. However, the mirror map of (\ref{eq:mir1}) is a global
identification over the whole parameter space and is less well
studied.  In \cite{MP:inst} this identification was proved for models
describing hypersurfaces in simplicial toric varieties.  For the more
general case considered here it is a natural conjecture.  In
particular, of course, this implies that mirror symmetry respects the
restriction of the full moduli spaces to the toric subspaces.  This is
known to be true when we have dual nef-partitions of reflexive
polytopes \cite{Bat:m,Boris:m}.

Our analysis above does not quite respect this mirror symmetry. The
GKZ determinant corresponding to decompactification of $x$ vacua is
not mirror to the condition corresponding to decompactification of
$\sigma$ vacua. In particular, we included points at vertices of
$\Conv\cB$ but we ignored the points at vertices of $\Conv\cA$.
We need to resolve this apparent violation of mirror symmetry.

The contribution of a vertex point in $\cB$ to $E_{\cB}$ is simply the
associated coefficient $b_\beta$ (perhaps raised to some power). So
the nonsingularity is associated to setting $b_\beta=0$. For mirror
symmetry to be respected, we need some kind of singularity when
$a_\alpha=0$, if $\alpha$ is a vertex point.

We can put a metric on $Z_\Sigma$ using the moment map in
$\frak{g}^*_{\R}$ which is given by the values of the
$\rho_\alpha$'s. Following \cite{AGM:mdmm} we have
\begin{prop}  \label{prop:pl}
  The particular triangulation we use for $\cA$ is the one consistent
  with a convex piecewise linear function taking the values of a
  ``height function'' $\log|a_\alpha|$ at each point $\alpha\in\cA$.
\end{prop}
Thus, the mirror to $b_\beta=0$ is setting the height function of
$\alpha$ to $-\infty$. This results in a component of $Z_\Sigma$
becoming infinitely large. Thus $X_\Sigma$ may decompactify.

The way the vertices of $\Conv\cA$ contribute to the discriminant is
therefore quite different to the other faces. The non-vertex faces are
associated to singularities induced by noncompactness of the Coulomb
branch while the vertices correspond to noncompactness of the Higgs
branch. For $\Conv\cB$, all the faces, whether vertices are not, give
singularities associated to noncompactness of the Higgs branch.

\subsection{Extremal Transitions and Breaking the Reflexive
  Condition} \label{ss:nonR}

Consider the left-most example in figure \ref{fig:Deg}. One might expect
to produce the discriminant of the familiar quadratic equation
$ax^2+bx+c$. But we get $ac(b^2-4ac)$ rather than
$b^2-4ac$. Essentially, this is
because $E_\beta$ looks for solutions of the weaker equations
$x_i\frac{\partial W}{\partial x_i}=0$ where $W=ax_0^2+bx_0x_1+cx_2^2$.
As such, $E_\beta=0$ is a stronger condition than non-compactness of 
$\Crit W$.
This means that the conditions we considered above may well be too
strong, and indeed they are. That is, in some cases we do not require
points in $\cB$ at the vertices of the dual cone of $\Conv\cA$.

Mirror to this should be the statement that we can sometimes drop
points at the vertices of $\Conv\cA$. Above we argued that dropping
such points amounts to a limit in K\"ahler moduli space where a
previously compact subset in $Z_\Sigma$ becomes infinitely
large. This need not cause problems with the \GLSM\ since $\Crit(W)$
may miss this infinitely large component.

It turns out that by considering extremal transitions we are forced to
consider non-reflexive pairs. First consider the following:
\begin{definition}
The pointsets $(\cA,\cB)$ are $\cB$-complete if
\begin{equation}
\Conv\bigl(\Cone(\Conv\cA)^\vee\cap M\cap H_\nu\bigr)
    = \Conv(\cB),
\end{equation}
and similarly $\cA$-complete if
\begin{equation}
\Conv\bigl(\Cone(\Conv\cB)^\vee\cap N\cap H_\mu\bigr)
    = \Conv(\cA),
\end{equation}
\end{definition}

Begin with pointsets $\cA$ and $\cB$
that correspond to a generically good theory (e.g., they are
reflexive). Now remove a point or points from $\cA$ to shrink the
convex hull. Thus $\Cone(\Conv\cA)^\vee$ grows. We can then add points
to $\cB$ so that $(\cA,\cB)$ is $\cB$-complete. This process
corresponds (if it is geometric) to an extremal transition as we now
review.

Obviously ignoring a point $\beta\in\cB$ amounts to setting
$b_\beta=0$. Thus, by mirror symmetry, we assume the same is true for
points in $\cA$. We therefore ignore $\alpha\in\cA$ by setting
$a_\alpha\to0$, i.e., the height function in proposition \ref{prop:pl}
to $-\infty$. This picks out a certain class of triangulations of
$\cA$. If the phase has a geometrical interpretation, it will typically
be an ``exoflop'' phase \cite{AdAs:masscat}. This has a line or
surface ``sticking out'' of a singular \CY\ component. We ignore the
massless D-branes by ignoring this extra component. Then we smooth the
remaining \CY\ component by a deformation of complex structure, by
adding to $W$ new allowed monomials, corresponding to the new points in
$\cB$, to complete the extremal transition.

Given our new enlarged pointset $\cB$ we should check to see if there
are any possible blow-ups by demanding $\cA$-completeness. So we start
with a model which is generically smooth and then go via an
extremal transition to a new theory. Since we have done all we can to
make this new theory smooth, as we argue shortly, it should be smooth!
This then gives the conjecture:
\begin{conjecture}  \label{conj:main}
If a \GLSM\ is defined by pointsets $\cA$ and $\cB$ which are both
$\cA$-complete and $\cB$-complete then this model is nonsingular for
generic values of the complex structure and K\"ahler parameters.
\end{conjecture}

\begin{figure}
\begin{center}
\begin{tikzpicture}[scale=1.0]
\draw (-0.5,1.8) node {$N_\R$};
\filldraw[gray!10] (0,0) -- (3,1.2) -- (3,-1.2);
\draw (0,0) -- (3,1.2);
\draw (0,0) -- (3,-1.2);
\draw (2,-1.5) -- (2,1.5);
\filldraw (2,-0.8) circle (0.07);
\filldraw (2,-0.4) circle (0.07);
\filldraw (2,0) circle (0.07);
\filldraw (2,0.4) circle (0.07);
\filldraw (2,0.8) circle (0.07);
\draw (2.5,0) node {$\cA$};
\draw (2,2) node {$\langle\mu,-\rangle=1$};
\draw[->] (1.7,1.7) .. controls (1.8,1.4) .. (2,1.3); 
\draw (0,-1) node {$\Cone(\Conv\cA)$};
\draw[->] (0,-0.6) -- (0.7,0);
\begin{scope}[xshift=7cm]
\draw (-0.5,1.8) node {$M_\R$};
\filldraw[gray!10] (0,0) -- (3,1.2) -- (3,-1.2);
\draw (0,0) -- (3,1.2);
\draw (0,0) -- (3,-1.2);
\draw (2,-1.5) -- (2,1.5);
\filldraw (2,-0.6) circle (0.07);
\filldraw (2,-0.3) circle (0.07);
\filldraw (2,0) circle (0.07);
\filldraw (2,0.3) circle (0.07);
\filldraw (2,0.6) circle (0.07);
\draw (2.5,0) node {$\cB$};
\draw (2,2) node {$\langle-,\nu\rangle=1$};
\draw[->] (1.7,1.7) .. controls (1.8,1.4) .. (2,1.3); 
\draw (0,-1) node {$\Cone(\Conv\cA)^\vee$};
\draw[->] (0.3,-0.6) -- (1.1,-0.4);
\draw[dotted] (0,0) -> (3,0.9);
\draw[dotted] (0,0) -> (3,-0.9);
\draw(4.6,0) node {$\left.\vphantom{\rule{1pt}{30pt}}\right\}\Cone(\Conv\cB)$};
\end{scope}
\end{tikzpicture}
\end{center} \caption{$\cB$-Completeness} \label{fig:dual}
\end{figure}
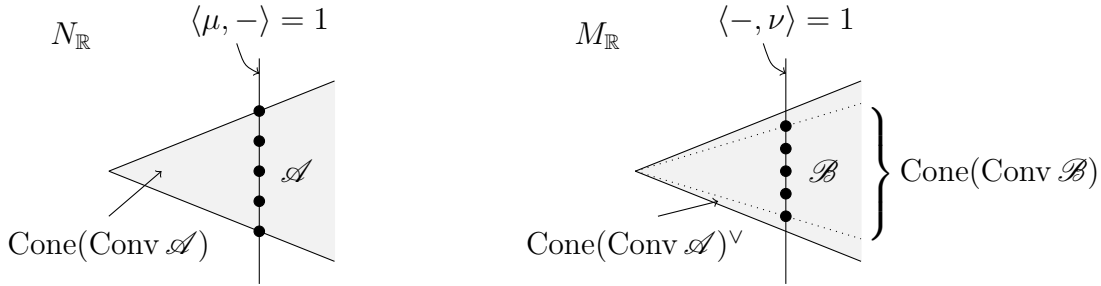

Note that if the pointsets are reflexive then they satisfy the
condition in the conjecture. We show schematically in figure
\ref{fig:dual} how one can satisfy $\cB$-completeness without
reflexivity.

From the perspective of string theory compactifications, a singular
conformal field theory corresponds to massless D-branes
\cite{Str:con}. Given such a string compactification one should be
free to deform it to give mass to the D-branes. In this sense, every
singular compactification should be infinitesimally close to a
nonsingular one. A counterexample to the above conjecture would,
na\"\i vely, contradict this idea. This counterexample, with generic
values for $a_\alpha$ and $b_\beta$ would require us to travel a
properly nonzero distance to undo the extremal transition to return to
a good theory.

Unfortunately we can't claim to have proven this conjecture by such an
argument. There are non-toric deformations that are not expressible in
terms of varying the parameters $a_\alpha$ and $b_\beta$. That is, our
\GLSM\ only parametrizes a ``toric'' sub-moduli space of the full
moduli space of theories. It may be possible that this toric sub-moduli
space lies within the codimension one subset of singular theories
within the moduli space. We have proven this never happens in the
reflexive case so it might be not unreasonable to expect the same in
general. This would justify the conjecture.

In section \ref{sss:BHnonR} we will see that even the above conjecture
is too strong. We will give an example of a model that is
$\cB$-complete but not $\cA$-complete and yet manages to be
generically nonsingular.


\section{Examples}  \label{s:ex}

\subsection{A Non-Reflexive Example} \label{sec:P34}

We can, at least, give an example of a non-reflexive \GLSM\ that is
generically nonsingular. Let $X$ be the intersection of a cubic and
quartic in weighted projective space
$\P^5_{\{2,1,1,1,1,1\}}$. According to the usual prescription
\cite{AG:gmi,Boris:m,MR2405763}, one uses the pointset $\cA$ given by
the rows of the matrix:
\begin{equation}
A^t = \left(\begin{matrix}
1&0&1&0&0&0&0\\
1&0&0&1&0&0&0\\
1&0&0&0&1&0&0\\
0&1&0&0&0&1&0\\
0&1&0&0&0&0&1\\
0&1&-2&-1&-1&-1&-1\\
1&0&0&0&0&0&0\\
0&1&0&0&0&0&0
\end{matrix}\right) \label{eq:P34A}
\end{equation}
These lie in a hyperplane given by $\mu=(1,1,0,0,0,0,0)$.
This example is not reflexive. The dual of the cone over $\Conv\cA$ is not
spanned by points living in a hyperplane. Equivalently, in the
language of \cite{Boris:m}, it is not a
nef partition of a reflexive polytope. However, we can define the
pointset $\cB$ from the rows of
\begin{equation}
B^t = \left(\begin{matrix}
0&1&0&0&0&2&-1\\
1&0&-1&-1&-1&4&0\\
1&0&-1&-1&-1&0&0\\
1&0&1&-1&-1&0&0\\
1&0&-1&3&-1&0&0\\
1&0&-1&-1&3&0&0\\
1&0&-1&-1&-1&0&4\\
0&1&0&0&0&-1&-1\\
0&1&1&1&0&-1&-1\\
0&1&1&0&1&-1&-1\\
0&1&1&0&0&0&-1\\
0&1&1&0&0&-1&0\\
0&1&1&0&0&-1&-1\\
0&1&0&3&0&-1&-1\\
0&1&0&0&3&-1&-1\\
0&1&0&0&0&-1&2
\end{matrix}\right)
\end{equation}
which lie in a hyperplane given by $\nu=(1,1,0,0,0,0,0)$. This does
satisfy the conditions of theorem \ref{th:main}.  Ignoring
coefficients, $A^tB$ defines the superpotential
\begin{multline}
W = x_8(x_2^3+x_3^3+x_4^3+x_5^3+x_6^3 + x_1x_2 +x_1x_3 +x_1x_4 +x_1x_5
+x_1x_6)\\ + x_7(x_1^2 + x_2^4 +x_3^4 + x_4^4 + x_5^4 + x_6^4), \label{eq:W34}
\end{multline}
as one would expect for this complete intersection. Note that
$\Conv\cB \cap M$ has 126 points. Including any of these points
maintains a good theory and corresponds to including further
``interior'' monomials in (\ref{eq:W34}). This yields a nonsingular
CFT since the intersection of a cubic and quartic in $\P^5_{211111}$
is generically smooth. Note that $\P^5_{211111}$ is not Gorenstein.

It is interesting ask whether $X$ has a geometric mirror. As we
discussed in section \ref{ss:geom}, this is not an easy question. Let
us look for a large radius phase for the mirror. That is, exchanging
the r\^oles of $\cA$ and $\cB$, can we find a phase of this model
corresponding to a complete intersection in a toric variety? We can
look for a special $(s-1)$-simplex. In our case we are looking for a
1-simplex with vertices on any of the 126 points in $\Conv\cB$ such
that each facet of $\Conv\cB$ contains precisely one vertex of this
simplex. In this, case a computer aided search shows that no such
1-simplex exists. Thus, there is no obvious ``geometric phase'', even
though with this many points the number of phases is incalculably
large. (The work of \cite{BFS:secpol} puts an upper bound of around
$10^{30000}$ on this number.)

\subsection{Extremal Transitions}  \label{sec:exm}

The motivation for conjecture \ref{conj:main} came from considering
extremal transitions. Extremal transitions are usually viewed from the
perspective of geometry but we can easily have transitions between
non-geometric phases too. We will give several examples of the
latter. This will also show how to transition between the
non-reflexive example of section \ref{sec:P34} and reflexive
cases. Thus one is forced to consider non-reflexive models if all
extremal transitions are followed.

\subsubsection{$\P^5|\,4\,\,2$}.

The intersection of a quartic and quadric in $\P^5$ is a \CY\
threefold and is given by reflexive data:

\begin{equation}
A^t = \left(\begin{matrix}
1&0&1&0&0&0&0\\
1&0&0&1&0&0&0\\
1&0&0&0&1&0&0\\
1&0&0&0&0&1&0\\
0&1&0&0&0&0&1\\
0&1&-1&-1&-1&-1&-1\\
1&0&0&0&0&0&0\\
0&1&0&0&0&0&0
\end{matrix}\right),
B^t = \left(\begin{matrix}
0&1&2&0&0&0&-1\\
1&0&3&-1&-1&-1&0\\
1&0&-1&-1&-1&-1&0\\
1&0&-1&3&-1&-1&0\\
1&0&-1&-1&3&-1&0\\
1&0&-1&-1&-1&3&0\\
1&0&-1&-1&-1&-1&4\\
0&1&0&0&0&0&-1\\
0&1&0&2&0&0&-1\\
0&1&0&0&2&0&-1\\
0&1&0&0&0&2&-1\\
0&1&0&0&0&0&1
\end{matrix}\right)
\end{equation}
\[
\mu=\nu=(1,1,0,0,0,0,0)
\]

$\cA$ has two triangulations. As promised, one gives $X$ as quadric
$\cap$ quartic in $\P^5$.  The other is a ``bad hybrid'' Landau--Ginzburg
fibration over $\P^1$.

Now do an extremal transition by dropping $(1,0,0,0,0,0,0)\in\cA$.  We
restore this to a reflexive pair if we add $(-1,2,1,1,1,1,-2)$ to
$\cB$. For this new theory the points in $\cA$ now form the vertices
of a simplex and so this model clearly has only one phase. This is a
\LG\ phase with superpotential
\begin{equation}
  W = x_1^4+x_2^4+x_3^4+x_4^4+x_5^4+x_6^4+bx_7^2+\ldots, \label{eq:LG1}
\end{equation}
where the extremal transitions ``joins on'' at $b=0$. That is, the
monomial $x_7^2$ is the extra point we add to $\cB$. Note that, as
usual, there is a finite group remnant of $G$ which gives us a \LG\
orbifold theory. In this case we have $\Z_4$ group acting.

Note that when $b\neq0$ we are allowed to ignore the massive
$x_7$. This yields the Gepner model $\mathbf{2}^6$. The resulting
model effectively has $h^{1,1}=0$ and is clearly not geometric.

It is interesting to picture physically what has happened in this
transition. We started with a geometry \CY\ threefold $X$ and then
deformed the K\"ahler form to shrink it down. The resulting type of
phase limit was analyzed in \cite{me:hybridm}. This limit is a hybrid
model over the weighted projective space $\P^1_{\{2,4\}}$ with fiber a
\LG\ model. The theory is at finite distance in moduli space and
singular. This ``badness'' of the hybrid model comes from the fact
that the base $\P^1_{\{2,4\}}$ is not fixed under the $\C^*$ action
from the $R$ symmetry.

This bad hybrid has 2 massless D-branes. The original \CY\ threefold
had $h^{1,1}=1$ and $h^{2,1}=89$. At this bad hybrid limit we turn on
a deformation of complex structure to smooth the theory and end up
with the Gepner model with $h^{1,1}=0$ and $h^{2,1}=90$. It is clear,
therefore, that this model is ``stuck'' at small radius when this 90th
deformation is turned on, and this is why there is no natural geometric
interpretation.

Having said that, there is a weaker geometrical interpretation in the
following sense. The D-brane category associated with the \LG\ model
(\ref{eq:LG1}) is the graded category of matrix factorizations of this
superpotential \cite{HW:mfac}. If we call this category $\cA$, by a
result of Orlov \cite{Orlov:mfc} we have a semiorthogonal decomposition
\begin{equation}
\DC(X_{\textrm{fake}}) =
  \langle \cA,\O,\O(1),\O(2),\O(3)\rangle,
\end{equation}
where $X_{\textrm{fake}}$ is a quartic hypersurface in $\P^6_{\{2,1,1,1,1,1,1\}}$.

The 5-fold $X_{\textrm{fake}}$ has a Hodge diamond
\setcounter{MaxMatrixCols}{12}
\begin{equation}
\begin{matrix}
&&&&&1\\
&&&&0&&0\\
&&&0&&1&&0\\
&&0&&0&&0&&0\\
&0&&0&&1&&0&&0\\
0&&1&&90&&90&&1&&0\\
&0&&0&&1&&0&&0\\
&&0&&0&&0&&0\\
&&&0&&1&&0\\
&&&&0&&0\\
&&&&&1
\end{matrix}
\end{equation}
The middle row is that of a \CY\ threefold with $h^{2,1}=90$ and
indeed we expect this to have the same complex structure moduli space
as the \LG\ theory. However, we wish to emphasize that the extremal
transition shows that the more natural physical interpretation is a
non-geometric phase {\em stuck at small radius\/} and not
$X_{\textrm{fake}}$.

It is perhaps worth emphasizing a construction which {\em doesn't\/}
work. One might be tempted to say the \LG\ orbifold associated with
the Gepner $\mathbf{2}^6$ theory has 6 fields and a superpotential
\begin{equation}
  W = x_1^4+x_2^4+x_3^4+x_4^4+x_5^4+x_6^4.  \label{eq:LGb}
\end{equation}
Correspondingly $X_{\mathrm{fake}}$ would be a quartic 4-fold in
$\P^5$. Again we have a category $\cA$ of matrix factorizations of
(\ref{eq:LGb}). The problem is that $\cA$ is not a \CY\ category. The
Serre functor $S_{\cA}$ obeys $S_{\cA}^2=[6]$ but not
$S_{\cA}=[3]$.\footnote{To see the former use section 4 of
  \cite{kuz:V14}. To see the latter use the action of $S_{\cA}$ on
  K-theory as in section 5.1.6 of \cite{AdAs:masscat}.}  In other
words, the open string spectrum does not exhibit spectral
flow. Similarly, the Hodge diamond of the quartic 4-fold is not at all
what we would want. Adding the $x_7^2$ term into the superpotential
fixes all this, and is what the extremal transition tells us to do.

\subsubsection{$\P^5_{\{2,1,1,1,1,1\}}|\,4\,\,3$}.

Now reconsider the intersection of a cubic and quadric in weighted 
$\P^5_{\{2,1,1,1,1,1\}}$ as in section \ref{sec:P34}. The pointset
$\cA$ given by the rows of (\ref{eq:P34A}) have two
triangulations. The first is the \CY\ phase as expected. The other is
another bad hybrid \LG\ fibration over $\P^1_{\{3,4\}}$. Just like
previous example this has a singular limit at a finite distance in
moduli space. We get an extremal transition by removing the point
$(1,0,0,0,0,0,0)\in\cA$. A generically smooth model is restored by
adding the point $(-3,4,3,3,3,-4,-4)$ to $\cB$.

The resulting theory has only a \LG\ phase with superpotential given
by
\begin{equation}
  W = x_1^4+x_2^4+x_3^4+x_4^4+x_5^4+bx_6^4+x_7^2+\ldots, \label{eq:LG2}
\end{equation}
where the extremal transitions ``joins on'' at $b=0$. That is, the
monomial $x_6^4$ is the extra point we add to $\cB$. But this is
exactly the same family as the previous example.

So the family of \LG\ theories of the form (\ref{eq:LG1}) (or
(\ref{eq:LG2})) contains the Gepner model $\mathbf{2}^6$ as well as
two limit points where the theory degenerates. These have extremal
transitions to the two \CY\ threefolds we have considered so far. In
particular, we have a path between the non-reflexive example of
section \ref{sec:P34} and the reflexive example of this section.

\subsubsection{$\P^6|\,3\,\,2\,\,2$}  

Since we have found an extremal transition to the Gepner model
$\mathbf{2}^6$, it is worth asking the question if we can find a
transition to the well-known, non-geometric Gepner model of
$\mathbf{1}^9$ which was the first Gepner model constructed \cite{Gep:}.

Indeed we can, and it turns out to be an example known to have a
singular limit as studied in \cite{me:hybridm}. The example consists
of the complete intersection of two quadrics and a cubic in $\P^6$.
So the superpotential is
\begin{equation}
W = x_8 f_2(x_1,\ldots,x_7) + x_9 g_2(x_1,\ldots,x_7) + x_{10}
h_3(x_1,\ldots,x_6),
\end{equation}
where the subscripts on the functions denote their homogeneous degree.

For this example the pointset $\cA$ has two triangulations again, with
one giving the interpretation as a \CY\ manifold with $h^{1,1}=1$ and
$h^{2,1}=73$. The other limit is a bad hybrid with base
$\P^2_{\{3,2,2\}}$. As discussed in \cite{me:hybridm}, this limit has
an $\SU(2)$ gauge symmetry that is inextricably linked to gravity and,
as such, has no field theory limit. Interestingly for us, there are
also 7 massless fundamentals of $\SU(2)$ in this limit
\cite{AA:toapp}. It follows
that there is a Higgs branch of the moduli space giving us an extremal
transition. 

Repeating the analysis of the previous sections to put this transition
into toric language, we lose point from $\cA$ corresponding to the
variable $x_{10}$ and we gain two points in $\cB$ corresponding to the
monomials $x_8^3$ and $x_9^3$. We have a superpotential which is
cubic in 9 variables, and a $\Z_3$-orbifold action from $G$. This is
the $\mathbf{1}^9$ model as promised.


\section{Ignoring Interior Points}  \label{s:int}

As we saw in the previous section, ignoring points at the vertices of
$\Conv\cA$ can result in singularities and thus extremal
transitions. In this section we consider ignoring points in $\cA$
which are not at the vertices. Thus we do not change $\Conv\cA$ and
thus the mode remains generically nonsingular. This has useful
applications in studying mirror pairs. In particular, the phase
picture becomes very daunting if there are many points in $\cA$. By
ignoring most of them we still get mirror pairs according to section
\ref{ss:mir} but the phase picture can be more manageable. 

An example of this idea which has been done many times implicitly
before is to drop points in codimension-one faces for Batyrev-type
mirror pairs. After all, it is known that such points do not contribute to
the Hodge numbers \cite{Bat:m}. We give two other examples of reducing
other types of mirror symmetry to that considered in this paper.

\subsection{Berglund--H\"ubsch Mirror Symmetry} \label{ss:BH} 

Mirror symmetry exchanges the matrices $A$ and $B$ and therefore takes
the transpose of the matrix of exponents in the superpotential
$P=A^tB$. In the special case that the model in question is a \LG\
theory, this should reduce to Berglund-H\"ubsch
\cite{Berglund:1991pp,Kraw:BH} mirror symmetry. We describe exactly
how this works in a particular example. All other cases are very
similar. 

Our observations here are certainly not the first concerning the
relationship of the Berglund-H\"ubsch construction to mirror symmetry
via toric geometry. See, for example, \cite{MR3046990,Cla:BH}. The
picture we give here has substantial overlap with \cite{FK:toricm} who
first realized that Berglund-H\"ubsch mirror symmetry arises from a
correct understanding of certain phases.

Suppose we have a smooth quintic hypersurface $X$ in $\P^4$ defined by
\begin{equation}
  x_1^4x_2 + x_2^4x_3 + x_3^4x_4 + x_4^4x_5 + x_5^5.  \label{eq:BH1}
\end{equation}
(This example first appeared in \cite{Greene:1991iv}.)
We completely understand mirror symmetry for quintics and so we should
certainly be able to construct the mirror. In particular, we have
the explicit mirror map (\ref{eq:mir1}) so we should know exactly
which K\"ahler data on the mirror corresponds to the choice of complex
structure (\ref{eq:BH1}). It follows that we should know which phase
the mirror is in.

\subsubsection{A Reflexive Model}   \label{sss:BHref}

The standard way to proceed is to use the well-known standard
reflexive pair of polytopes for the quintic and its mirror. The data
for the mirror is
\begin{equation}
A^t=
\begin{pmatrix}1&4&-1&-1&-1\\1&-1&4&-1&-1\\
1&-1&-1&4&-1\\1&-1&-1&-1&4\\1&-1&-1&-1&-1\\&&\vdots\end{pmatrix},
B^t=
\begin{pmatrix}1&0&0&0&0\\1&1&0&0&0\\1&0&1&0&0\\
1&0&0&1&0\\1&0&0&0&1\\1&-1&-1&-1&-1\end{pmatrix},
\end{equation}

The pointset $\cA$ has 126 points and so we only show those at the
vertices of convex hull above. With this many points, an analysis of
the phase picture is hopeless. What we do instead is choose $\cA$ to
be the 5 points which are the vertices of $\Conv\cA$ together with 4
more points given by the first 4 monomials in (\ref{eq:BH1}). We are
therefore considering mirror of the family of quintics whose defining
equation has 9 monomials --- the union of the Fermat ones and those in
(\ref{eq:BH1}). The resulting pointset has a more manageable 42
triangulations.  One obvious triangulation consists of the single
simplex given by the convex hull. This is mirror to the familiar \LG\
theory $\C^5/(\Z_5^{\oplus4})$.

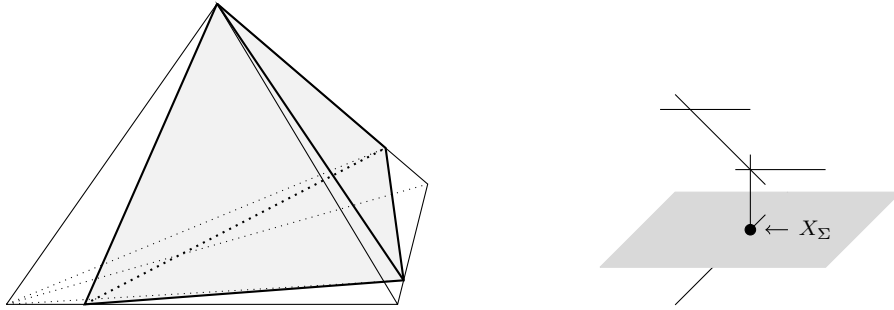
\begin{figure}
\begin{center}
\begin{tikzpicture}[scale=0.8]
\coordinate (x5) at (4.5,5);
\coordinate (x4) at (8,2);
\coordinate (x3) at (7.5,0);
\coordinate (x2) at (1,0);
\coordinate (a5) at ($ (x2) !.2! (x3) $);
\coordinate (a7) at ($ (x3) !.2! (x4) $);
\coordinate (a8) at ($ (x4) !.2! (x5) $);
\filldraw[gray!10] (a5) -- (a7) -- (a8) -- (x5) -- (a5);
\draw (x2) -- (x3) -- (x4) -- (x5) -- (x3);
\draw (x5) -- (x2);
\draw[dotted] (x2) -- (x4);
\draw[thick] (x5) -- (a5) -- (a7) -- (a8) -- (x5);
\draw[thick,dotted] (a5) -- (a8);
\draw[thick] (x5) -- (a7);
\draw[dotted] (x2) -- (a8);
\draw[dotted] (x2) -- (a7);
\end{tikzpicture}
\hspace{20mm}
\begin{tikzpicture}[scale=1]
\coordinate (a) at (3,3);
\draw (3.5,3.5) -- (2,2);
\filldraw[gray!30] (1,2.5) -- (4,2.5) -- (5,3.5) -- (2,3.5) -- (1,2.5);
\draw (3.2,3.2) -- (a);
\draw (a) -- (3,4);
\draw (2.8,3.8) -- (4,3.8);
\draw (3.2,3.6) -- (2,4.8);
\draw (1.8,4.6) -- (3,4.6);
\filldraw (a) circle (0.07);
\draw[<-] (3.2,3) -- (3.5,3) node[anchor=west] {$\scriptstyle X_\Sigma$};
\end{tikzpicture}
\end{center}
\caption{A Hidden \LG\ Theory.} \label{fig:PLG}
\end{figure}

Four of the triangulations contain a simplex with vertices
corresponding to the monomials in (\ref{eq:BH1}). Each of these is
consistent with varying the complex structure so that we return to the
defining equation (\ref{eq:BH1}).  We show such a triangulation
schematically (in one fewer dimensions) on the left in figure
\ref{fig:PLG}. Now consider the toric variety $Z_\Sigma$ associated to
one such triangulation. It is certainly not of the form $\C^5/H$, for
some finite group $H$, as one would normally expect for a \LG\
phase. Instead the set of compact $T_N$-orbits comprise a surface and
set of $\P^1$'s as shown on the right in figure
\ref{fig:PLG}. However, when one computes the critical points of $W$
to compute $X_\Sigma$, one finds that this is a single point. So the
theory really is a \LG\ orbifold theory. Usually one associates a \LG\
phase with a triangulation consisting of a single simplex, so this
might be regarded as a ``hidden'' \LG\ phase.

Locally, at the point $X_\Sigma$, the geometry of $V_\Sigma$ is
$\C^5/\Z_{256}$ with group generator acting with weights
$\ff1{256}(1,251,20,176,64)$. We have therefore shown that the mirror
of the quintic hypersurface with defining equation (\ref{eq:BH1}) is a
\LG\ orbifold of the form $\C^5/\Z_{256}$. The superpotential is the
transposed form of (\ref{eq:BH1}):
\begin{equation}
x_1^5x_2 + x_2^4x_3 + x_3^4x_4 + x_4^4x_5 +x_5^4.
\end{equation}

We can produce a geometric mirror by deforming the complex structure
of the quintic by adding the monomial $x_1x_2x_3x_4x_5$ to
(\ref{eq:BH1}) with a large coefficient.  In the mirror, on the
K\"ahler side, this amounts to including the point $(1,0,0,0,0)$ in
the triangulation. This blows up the $\C^5/\Z_{256}$ singularity and
$X_\Sigma$ becomes a degree 256 hypersurface in
$\P^4_{\{41,48,51,52,64\}}$.

We claim that all Berglund--H\"ubsch examples arise this way. In
particular any defining equation with only 5 terms that leads to a
smooth \CY\ will be mirror to a \LG\ orbifold theory. The five
monomials become vertices of a 4-simplex in $\cA$ and the smoothness
condition guarantees that the critical pointset of the superpotential
is an isolated point.

\subsubsection{A Non-Reflexive Version} \label{sss:BHnonR}

Now let's do the same analysis in a more direct way. Simply set $\cA$
to only 5 points given by the monomials in (\ref{eq:BH1}). That is
\begin{equation}
A^t = \begin{pmatrix}1&-1&-1&-1&-1\\
1&3&0&-1&-1\\1&-1&3&0&-1\\
1&-1&-1&3&0\\1&-1&-1&-1&3\end{pmatrix}
\end{equation}
The dual of the cone over $\Conv\cA$ is then generated by the rows of
the following matrix:
\begin{equation}
\begin{pmatrix}1&-51/41&-52/41&-48/41&-64/41\\
1&1&0&0&0\\
1&-1/3&4/3&0&0\\
1&1/13&-4/13&16/13&0\\
1&-1/51&4/51&-16/51&64/51\end{pmatrix}
\end{equation}
Now intersect this cone with the lattice $M$ and the hyperplane
$H_\nu$, where $\nu=(1,0,0,0,0)$. The result is the pointset $\cB$
given by rows of the matrix
\begin{equation}
B^t = \begin{pmatrix}1&0&0&0&0\\1&1&0&0&0\\1&0&1&0&0\\
1&0&0&1&0\\1&0&0&0&1\\1&-1&-1&-1&-1\end{pmatrix},
\end{equation}
which is, of course, the data for the quintic. Thus we recover the
quintic as above. We only have one phase, a \LG\ phase on
$\C^5/\Z_{256}$ as above. Note that we do not have any of the
irrelevant detritus sticking out of the \LG\ theory as we had in
figure \ref{fig:PLG}.

This is a perfectly good \GLSM. It is $\cB$-complete but not
$\cA$-complete. So the conditions in conjecture \ref{conj:main} do not
give {\em necessary\/} conditions for a generically nonsingular family
of models.

\subsection{Vafa--Witten Mirror Symmetry} \label{ss:VW}

Perhaps a little gratuitously, let us give another example of dropping
internal points to simplify mirror symmetry. In \cite{VW:tor} a mirror
pair of orbifolds were constructed using discrete torsion. Let $E$ be
an elliptic curve and let $X^\sharp$ be the orbifold $(E\times E\times
E)/H$, where $H=\Z_2\times\Z_2$ generated by $(-1,-1,1)$ and
$(1,-1,-1)$. $X^\sharp$ may be blown-up to a smooth \CY\ threefold $X$
with $h^{1,1}=51$ and $h^{2,1}=3$.

Since $H^2(H,\GU(1))=\Z_2$, we may consider the orbifold with discrete
torsion switched on. It \cite{VW:tor} it was argued that the resulting
theory is mirror to the orbifold without discrete torsion. It was also
argued that a generic member of the resulting family, $Y$, mirror to
$X$, has 64 ``stuck'' nodes. This was also analyzed in \cite{AMG:stab}
in terms of the Brauer group. This Brauer group interpretation was
extended to other models in \cite{Caldararu:2007tc,Add:P3} in a way
relevant for us here.

Let the elliptic $E$ be realized as a quartic $x_1^2+x_2^4+x_3^4$ in
$\P^2_{\{2,1,1\}}$. Then ``$-1$'' acts on $E$ as
$[x_1,x_2,x_3]\mapsto[-x_1,x_2,x_3]$. In our desired \GLSM\ language
we will write this as the set of points
\begin{equation}
A^t = \begin{pmatrix}1&0&0\\1&1&0\\1&0&1\\1&-2&-1\end{pmatrix}. \label{eq:E}
\end{equation}
The ``$-1$'' is part of the torus $T_N$ and thus can also be written
as a vector in $N\otimes\Q$. This vector is $(\ff12,\ff12,0)$.

To build $E\times E\times E$ we simply take the direct sum of three
copies of (\ref{eq:E}) so that $N$ is now 9-dimensional. The action of
$H$ is then generated by
\begin{equation}
\begin{split}
&(\ff12,\ff12,0,\ff12,\ff12,0,0,0,0)\\
&(0,0,0,\ff12,\ff12,0,\ff12,\ff12,0).
\end{split}  \label{eq:Z2Z2}
\end{equation}
To perform the orbifold in toric language, we fix the pointset $\cA$
but modify the lattice $N$ by refining it minimally to include the two
vectors (\ref{eq:Z2Z2}). One then transforms the matrix $A$ according
to a change of basis to one that generates the new $N$.

This results in a reflexive pair of points $\cA$ and $\cB$ so we
should know exactly what the mirror of $X$ is. The convex hull of both
$\cA$ and $\cB$ are simplices with 9 vertices. That means they both
have a phase that is purely a \LG\ orbifold. Indeed, they are both
orbifolds of the Gepner model $\mathbf{2}^6$ and so this mirror
symmetry is actually of the Greene--Plesser type \cite{GP:orb}.

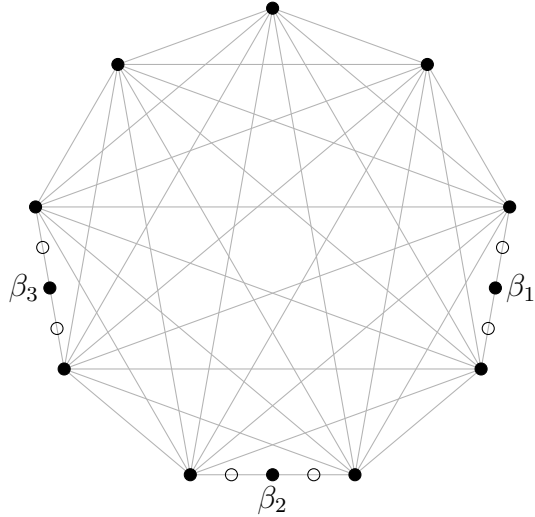
\begin{figure}
\begin{center}
\begin{tikzpicture}[scale=0.8]
\foreach \x in {0,...,8} {
  \coordinate (c\x) at ({4*sin(\x*360/9)},{4*cos(\x*360/9)});
  \foreach \y in {0,...,\x}
    \draw[gray!60] (c\x) -- (c\y);
}  
\foreach \x in {0,...,8}
  \filldraw (c\x) circle (0.1);
\filldraw ($ (c2) !.5! (c3) $)  circle (0.1) node[anchor=west] {$\beta_1$};
\filldraw ($ (c4) !.5! (c5) $)  circle (0.1) node[anchor=north] {$\beta_2$};
\filldraw ($ (c6) !.5! (c7) $)  circle (0.1) node[anchor=east] {$\beta_3$};
\draw ($ (c2) !.25! (c3) $)  circle (0.1);
\draw ($ (c2) !.75! (c3) $)  circle (0.1);
\draw ($ (c4) !.25! (c5) $)  circle (0.1);
\draw ($ (c4) !.75! (c5) $)  circle (0.1);
\draw ($ (c6) !.25! (c7) $)  circle (0.1);
\draw ($ (c6) !.75! (c7) $)  circle (0.1);
\end{tikzpicture}
\end{center}
\caption{The Pointset $\cB$ for the orbifold $T^6/\Z_2\times\Z_2$.}
   \label{fig:VW}
\end{figure}

The pointset $\cB$ in this 8-simplex is shown in figure
\ref{fig:VW}. There are 18 points with 9 at the vertices of the
simplex. From now on we wish to describe the mirror of $X$ and thus we
exchange the roles of $\cA$ and $\cB$. Points in figure \ref{fig:VW} now
correspond to homogeneous coordinates.

We will choose to ignore the 6 points shown as unfilled
circles. The remaining 12 points have 8 triangulations. These 8
triangulations are easy to understand; one can either include or omit
from the triangulation each of the 3 points shown as $\beta_1$,
$\beta_2$, $\beta_3$ in the figure.

The superpotential in these 12 variables can be written
\begin{equation}
  W = \begin{pmatrix}x_1&x_2&x_3\end{pmatrix} M
      \begin{pmatrix}x_1\\x_2\\x_3\end{pmatrix}
  + x_{10}^2 + x_{11}^2 + x_{12}^2,
\end{equation}
where
\begin{equation}
  M = \begin{pmatrix}f_{400}&f_{220}&f_{202}\\
     f_{220}&f_{040}&f_{022}\\ 
     f_{202}&f_{022}&f_{004}\end{pmatrix},
\end{equation}
and $f_{ijk}$ is a homogeneous function of degree $(i,j,k)$ in the
variables $(x_4,x_5)$, $(x_6,x_7)$, $(x_8, x_9)$ respectively. The
notation is chosen to coincide with that of \cite{AMG:stab}.

In the triangulation where we include all three points $\beta_1$,
$\beta_2$, $\beta_3$, $V_\Sigma$ is a vector bundle over
$\P^1[x_4,x_5]\times\P^1[x_6,x_7]\times\P^1[x_8,x_9]$. We therefore
have a hybrid model with base $\P^1\times\P^1\times\P^1$ and fibre
given by a quadratic \LG\ theory. This situation is essentially
identical to the model of 4 quadrics in $\P^7$ which has been
extensively analyzed in
\cite{AG:gmi,Add:P3,MR2419925,Caldararu:2007tc}.  Accordingly, we
interpret this hybrid model as a double cover of
$\P^1\times\P^1\times\P^1$ branched over the surface given by $\det
M=0$. This surface has 64 nodes as does the double cover.  A Brauer
class is ``switched on'' in the same way as it occurs in
\cite{Add:P3}.  This precisely agrees with Vafa and Witten's mirror
\cite{VW:tor}.

It is worth mentioning the other phases. Consider the phase where we
include $\beta_1$ but not $\beta_2$ or $\beta_3$ in the triangulation.
This is a hybrid model with base $\P^1[x_4,x_5]$. To see the fibre,
fix the values of $x_4$ and $x_5$ which gives a \LG\ superpotential
quartic in $x_6,x_7,x_8,x_9$ and quadratic in
$x_1,x_2,x_3,x_{10},x_{11},x_{12}$. The usual arguments go that we can
ignore the quadratic terms in a superpotential\footnote{To get the
  open string spectrum correct we use Kn\"orrer periodicity, which
  says we can ignore an {\em even\/} number of variables appearing
  quadratically.\cite{Orlov:mfc}} leaving us with a quartic in four
variables. This, of course is equivalent to a quartic K3 surface and
thus we {\em appear\/} to have a K3-fibration.

The problem is that there are 4 points on the base $\P^1$ where the
rank of $M$ drops and the variables we were ignoring have a vanishing
coefficient in their quadratic term. Thus, at the bad fibres, it is
these ignored variables that do something interesting, and not the K3
part. It follows that we do not have a family of K3's near the bad
fibres, and so we do not have a K3 fibration. Similarly we have a
broken elliptic fibration over $\P^1\times\P^1$ in some of
the other phases.

\section*{Discussion} \label{sec:conc}

Clearly it would be nice to find a proof of, or counterexample to, our
conjectured sufficient condition for a generically nonsingular
model. An obvious way to find a counterexample is to search examples
via a random walk through extremal transitions. Indeed, the
formulation of the conjecture is well-suited to such a search.
It would also be very nice to find a necessary condition rather than a
sufficient condition. We should note that when $\Conv\cA$ is a
simplex, the work of \cite{Kreuzer:1992bi} effectively solves this
problem. The more general situation is, presumably, even more complicated.

If the conjecture is true, a search may generate many more examples of
nonreflexive mirror pairs. The famous list \cite{Kreuzer:2000xy} of
reflexive pairs for the hypersurface case (i.e.,
$\langle\mu,\nu\rangle=1$) runs to half a billion. Progress beyond
this special case has been limited, but has generally involved notions
of ``nef partitions'' following \cite{Boris:m}. If one simply wants to
enumerate possible string compactifications or \CY\ categories then
this notion is irrelevant. A full enumeration might be simply too
large to practically achieve, but it would be interesting to see if
certain properties are common. For example, how likely is it that any
geometric phase exists for a given model?  A large-radius limit may,
or may not, arise as a geometric phase limit as discussed in \ref{ss:geom}, so a
separate question would be how likely is it that a large-radius limit
exists for a given model.

The simple way in which (\ref{eq:mir1}) implements mirror symmetry is
in a sense surprising.  There is, for example, no {\em a priori\/}
reason from the point of view of \SCFT\ why mirror symmetry should
preserve the toric subspace.  This suggestive structure led to the
idea that mirror symmetry can be simply implemented as a T-duality
transformation directly in the \GLSM\
\cite{Morrison:1995yh,Hori:2000kt}.   This duality would imply that
if one can write a condition on the combinatorial data that is both
necessary and sufficient for the \GLSM\ to be generically nonsingular
it too should be mirror symmetric.  Finding this is
not likely to be straightforward, since in particular it would involve
refining the GKZ determinant to a necessary and sufficient condition,
but examples like \ref{sss:BHnonR} may be helpful as nontrivial test
cases. 

Our construction of the GKZ determinant (\ref{eq:GKZd}) from the
\GLSM\ is not quite complete. We have not derived the powers
$u(\Gamma)$. A more complete analysis of the model based on $\cA_h$ in
section \ref{sss:Afaces} should surely explain these numbers. This
should be investigated further.

\section*{Acknowledgments}

We thank Nick Addington, David Favero, Ilarion Melnikov and David
Morrison for helpful discussions. PSA is supported by NSF grants
DMS--1207708. MRP is supported by NSF grant PHY--1217109.  Any
opinions, findings, and conclusions or recommendations expressed in
this material are those of the authors and do not necessarily reflect
the views of the National Science Foundation.


\end{document}